\documentclass[11pt]{article}
\usepackage{graphicx}
\usepackage{subfigure}

\newcommand{\BABARPubYear}    {04}

\newcommand{\BABARConfNumber} {26}
\newcommand{\SLACPubNumber} {10611}

\input pubboard/babarsym

\long\def\inst#1{\par\nobreak\kern 4pt\nobreak
    {\it #1}\par\vskip 10pt plus 3pt minus 3pt}

\def\btn {\ensuremath{B^{+} \to \tau^{+} \nu_{\tau}}}
\def\btodlnux {\ensuremath{\Bub \to \Dz \ell^{-} \bar{\nu}_{\ell} X}}

\def\btodszlnu {\ensuremath {\B^- \to D^{*0} \ell^- \nulb}}
\def\dszlnu {\ensuremath {D^{*0} \ell^- \nulb}}
\def\tautoenunu {\ensuremath {\tau^+ \to e^+ \nu_e \nutb}}
\def\enunu {\ensuremath {e^+ \nu_e \nutb}}
\def\tautomununu {\ensuremath {\tau^+ \to \mu^+ \nu_{\mu} \nutb}}
\def\mununu {\ensuremath {\mu^+ \nu_{\mu} \nutb}}
\def\tautopinu {\ensuremath {\tau^+ \to \pi^+ \nutb}}
\def\pinu {\ensuremath {\pi^+ \nutb}} 
\def\tautopipiznu {\ensuremath {\tau^+ \to \pi^+ \pi^{0} \nutb}}
\def\pipiznu {\ensuremath {\pi^+ \pi^{0} \nutb}}
\def\tautothreepinu {\ensuremath {\tau^+ \to \pi^+ \pi^{-} \pi^{+} \nutb}}
\def\threepinu {\ensuremath {\pi^+ \pi^{-} \pi^{+} \nutb}}

\def\onlumi    {\ensuremath { 112.5  \invfb\ }}
\def\offlumi   {\ensuremath { 11.9 \invfb\  }}

\def\nBB   {\ensuremath {124.1 \times 10^{6}} }

\def\etal {{\it et al.}}

\begin{document}
{\pagestyle{empty}

\begin{flushright}
\babar-CONF-\BABARPubYear/\BABARConfNumber \\
SLAC-PUB-\SLACPubNumber \\
July 2004 \\
\end{flushright}

\par\vskip 5cm

\begin{center}
\Large \bf A Search for \btn \xspace Recoiling Against \btodszlnu
\end{center}
\bigskip

\begin{center}
\large The \babar\ Collaboration\\
\mbox{ }\\
\today
\end{center}
\bigskip \bigskip

\begin{center}
\large \bf Abstract
\end{center}

We present a search for the decay \btn\ in \nBB \xspace $\FourS$
decays recorded with the \babar\ detector at the SLAC PEP-II $B$-Factory. 
A sample of events with one reconstructed 
exclusive semi-leptonic $B$ decay (\btodszlnu) is selected, and
in the recoil a search for $\btn$ signal is performed. The $\tau$ is
identified in the following channels: $\tautoenunu$, $\tautomununu$,
$\tautopinu$, $\tautopipiznu$, $\tautothreepinu$. We find no evidence 
of signal, and we set a preliminary upper limit on the branching
fraction of $\mathcal{B}(\btn) < 4.3 \times 10^{-4}$ at the 90\%
confidence level (CL). This result is then combined with a statistically
independent \babar\ search for \btn\ to give a combined preliminary limit of
$\mathcal{B}(\btn) < 3.3 \times 10^{-4}$ at 90\% CL.

\vfill
\begin{center}

Submitted to the 32$^{\rm nd}$ International Conference on High-Energy Physics, ICHEP 04,\\
16 August---22 August 2004, Beijing, China

\end{center}

\vspace{1.0cm}
\begin{center}
{\em Stanford Linear Accelerator Center, Stanford University, 
Stanford, CA 94309} \\ \vspace{0.1cm}\hrule\vspace{0.1cm}
Work supported in part by Department of Energy contract DE-AC03-76SF00515.
\end{center}

\newpage
} 

\begin{center}
\small

The \babar\ Collaboration,
\bigskip

%
B.~Aubert,
R.~Barate,
D.~Boutigny,
F.~Couderc,
J.-M.~Gaillard,
A.~Hicheur,
Y.~Karyotakis,
J.~P.~Lees,
V.~Tisserand,
A.~Zghiche
\inst{Laboratoire de Physique des Particules, F-74941 Annecy-le-Vieux, France }
% ...

\end{center}\newpage

\section{INTRODUCTION}
\label{sec:Introduction}

\begin{figure}[htb]
\begin{center}
\includegraphics[width=0.45\textwidth]{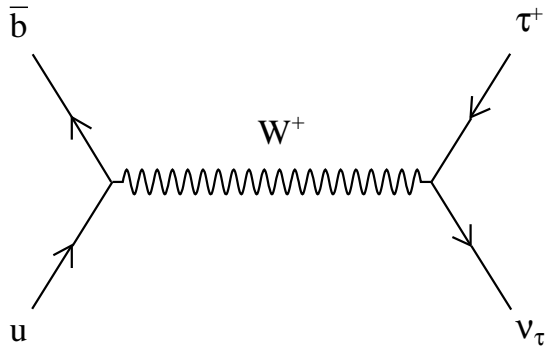}
\end{center}
\caption{\label{fig:feynman_diagram}%
Purely leptonic $B$ decay $\Bu \to \tau^{+} \nu_{\tau}$  
proceeds via quark annihilation into a $W^{+}$ boson.}
\end{figure}

In the Standard Model (SM), the purely leptonic decay \btn\ 
\footnote{Charge-conjugate modes are implied throughout this paper. The signal $B$ will always be denoted as a \Bu\ decay while the semi-leptonic $B$ will be denoted as a \Bub\ to avoid confusion.}
proceeds via quark annihilation 
into a $W^{+}$ boson (Fig. \ref{fig:feynman_diagram}).
Its amplitude is thus proportional to the product of the 
$B$-decay constant $f_B$ and the quark-mixing-matrix 
element $V_{ub}$. The branching fraction is given by:
\begin{equation}
\label{eqn:br}
\mathcal{B}(B^{+} \rightarrow {\taup} \nu)= 
\frac{G_{F}^{2} m^{}_{B}  m_{\tau}^{2}}{8\pi}
\left[1 - \frac{m_{\tau}^{2}}{m_{B}^{2}}\right]^{2} 
\tau_{\Bu} f_{B}^{2} \mid V_{ub} \mid^{2},\label{eq:brsm} 
\end{equation}
where we have set $\hbar = c = 1$, %
$G_F$ is the Fermi constant, 
$V_{ub}$ is a quark mixing matrix element~\cite{ref:c,ref:km}, 
$f_{B}$ is the $\Bu$ meson decay constant which describes the
overlap of the quark wave-functions inside the meson,
$\tau_{\Bu}$ is the $\Bu$ lifetime, and
$m^{}_{B}$ and $m_{\tau}$ are the $\Bu$ meson and $\tau$ masses.
This expression is entirely analogous to that for pion decay.
Physics beyond the SM, such as supersymmetry or two-Higgs doublet models,
could enhance $\mathcal{B}(\btn)$ by up to a factor of five through the
introduction of a charged Higgs boson~\cite{bib:higgs}.

Current theoretical values for $f_B$ 
(obtained from lattice QCD calculations)~\cite{ref:pdg2004} have  
large uncertainty, and purely leptonic decays of the $\Bu$ meson
may be the only clean experimental method of measuring
$f_B$ precisely. Given measurements of $|V_{ub}|$ from semi-leptonic
processes such as $B \to \pi \ell \nu$, $f_{B}$ could be extracted
from the measurement of the \btn\ branching fraction. In addition,
by combining the branching fraction measurement with results
from $B$ mixing, the ratio $|V_{ub}|/|V_{td}|$ can be extracted
from $\mathcal{B}(\btn)/\Delta m$, where $\Delta m$ is the
mass difference between the heavy and light neutral $B$ meson states.

The decay amplitude is proportional to the lepton mass and
decay to the lighter leptons is severely suppressed. This mode
is therefore the most promising for discovery at existing experiments.
However, challenges 
such as the large missing momentum from several neutrinos make 
the signature for \btn\ less distinctive than for other leptonic modes.

The SM estimate of this branching fraction
is $(9.3 \pm 3.9) \times 10^{-5}$, 
using $\Vub = (3.67 \pm 0.47)\times 10^{-4}$ and $f_{B} =
0.196 \pm 0.032$~\cite{ref:pdg2004} in Eq.~\ref{eq:brsm}.

Purely leptonic $B$ decays have not yet been observed. CLEO~\cite{cleo} and experiments 
at LEP~\cite{l3,aleph,delphi} have searched for this process and set limits on the 
branching fraction at the 90\% CL.
The most stringent limit on $\mathcal{B}(\btn)$ comes from the
L3 experiment~\cite{l3}:
\begin{eqnarray}
\mathcal{B}(\btn) & < & 5.7 \times 10^{-4} \, \textrm{at the 90\% CL. }
\end{eqnarray}
The \babar\ collaboration has results on the search for \btn\ 
decays~\cite{babar_prl_btn} using a sample of 
$88.9 \times 10^6$ $\FourS$ decays; the upper limit is:
\begin{eqnarray}
\mathcal{B}(\btn) & < & 4.2 \times 10^{-4} \, \textrm{at the 90\% CL. }
\end{eqnarray}
In this paper we report on a preliminary result from a different analysis
on a larger dataset.

\section{THE \babar\ DETECTOR AND DATASET}
\label{sec:babar}
The data used in this analysis were collected with the \babar\ detector
at the \pep2\ storage ring. 
The sample corresponds to an integrated
luminosity of \onlumi \xspace at the \FourS\ resonance (on-resonance) 
and \offlumi \xspace taken $40\mev$ below the \FourS\ resonance 
(off-resonance). The on-resonance sample consists of
about $(124.1 \pm 1.4) \times 10^{6}$  $\FourS$ decays (\BB\ pairs). The collider is operated with asymmetric
beam energies, producing a boost of $\beta\gamma \approx 0.56$ 
of the \FourS along the collision axis.

The \babar\ detector is optimized for asymmetric energy collisions at a
center-of-mass (CM) energy corresponding to the \FourS\ resonance.
The detector is described in detail in reference~\cite{ref:babar}. 
Charged particle tracking is provided by a five-layer double-sided silicon 
vertex tracker (SVT) and a 40-layer drift chamber (DCH) contained 
within the magnetic field of a 1.5 T superconducting solenoid.
The tracking system provides momentum reconstruction of charged
particles and measures energy loss ($dE/dx$) for particle identification.
Additional charged $K$--$\pi$ particle identification is provided
by a ring-imaging Cherenkov detector (DIRC), which exploits the 
total internal reflection of Cherenkov photons within the 
purified quartz bars. The energies of neutral particles are 
measured by an electromagnetic calorimeter (EMC) composed of 
6580 CsI(Tl) crystals, which provides an energy resolution 
$\sigma_{E}/E = (2.3 / E^{1/4} \oplus 1.9 ) \%$, 
where E is in \gev. For 20 \mev\ and 1 \gev\ clusters the 
EMC energy resolutions are $\sim$6\% and $\sim$3\%, respectively.
The magnetic flux return of
the solenoid (IFR) is instrumented with resistive plate chambers in order
to provide muon and neutral hadron identification.

A GEANT4-based \cite{geant4} Monte Carlo (MC) 
simulation is used to model the signal
efficiency and the physics backgrounds. Simulation samples
equivalent to approximately three times the accumulated data  were
used to model \BB\ events, and samples equivalent to approximately
1.5 times the accumulated data were used to model $\epem \to$
\uubar, \ddbar, \ssbar, \ccbar, and \tautau\ events. 
A large sample of signal events is simulated, where
one of the $B$ meson decays to $\btn$. 
Beam related background and detector noise 
from data are overlaid on the simulated events.

\section{ANALYSIS METHOD}
\label{sec:Analysis}

Due to the presence of multiple neutrinos, the \btn \xspace decay mode
lacks the kinematic constraints which are usually exploited in $B$ decay 
searches in order to reject both continuum and $B\overline{B}$ backgrounds.
The strategy adopted for this analysis is to reconstruct exclusively 
the decay of one of the $B$ mesons in the event, referred to as ``tag'' $B$, 
and to compare the remaining particle(s) in the event, referred as the 
``signal side'', with the signature
expected for the decay \btn. In order to avoid experimenter bias, the 
signal region in data is not examined (``blinded'') until the selection is
optimized based on MC simulation.

The tag $B$ is reconstructed in the set of decay modes \btodszlnu, where 
$\ell$ is $e$ or $\mu$. The $D^{*0}$ is reconstructed in $D^{0}\piz$ and
$\Dz \gamma$ modes. The $\Dz$ is reconstructed in four decay modes:
$K^{-}\pi^{+}$, $K^{-}\pi^{+}\pi^{-}\pi^{+}$, $K^{-}\pi^{+}\pi^{0}$, and
$K_{s}^{0}\pi^{+}\pi^{-}$. The $K_{s}^{0}$ is reconstructed only in the
mode $K_{s}^{0} \rightarrow \pi^{+}\pi^{-}$.
On the signal side the \btn \xspace signal is searched for in
both leptonic and hadronic $\tau$ decay modes:
$\tautoenunu$, $\tautomununu$, $\tautopinu$, $\tautopipiznu$,
$\tautothreepinu$. The branching fractions of the above $\tau$ decay 
modes are listed in Table \ref{tab:TauDecayModes}. 
Most of the kinematic variables used for 
event selection and background rejection are measured in the
CM frame.

\begin{table}[h]
\caption{\label{tab:TauDecayModes} Branching fractions for the $\tau$ decay modes used in the \btn\ search~\cite{ref:pdg2004}.}
\begin{center}
\begin{tabular}{|l|c|}
\hline
Decay Mode   &   Branching Fraction (\%)  \\
\hline\hline
$\tautoenunu$      & 17.84 $\pm$ 0.06 \\ \hline
$\tautomununu$     & 17.36 $\pm$ 0.06 \\ \hline
$\tautopinu$       & 11.06 $\pm$ 0.11 \\ \hline
$\tautopipiznu$    & 25.42 $\pm$ 0.14 \\ \hline
$\tautothreepinu$  & 9.16 $\pm$  0.10 \\ \hline
\end{tabular}
\end{center}
\end{table}

\subsection{TAG $B$ RECONSTRUCTION}
\label{sec:TagReco}

The tag $B$ reconstruction proceeds as follows. First we reconstruct the 
$\Dz$ candidates in the above four decay modes using tracks
and/or a $\piz$. The tracks
are required to meet particle identification criteria consistent
with the particle hypothesis, and are required to converge at a common vertex.
The $\piz$ candidate is required to have invariant mass between 
0.115--0.150 \gev/$c^2$ and its daughter photon candidates must 
have a minimum energy of 30 \mev.
The mass of the reconstructed $\Dz$ candidates in 
$K^{-}\pi^{+}$, $K^{-}\pi^{+}\pi^{-}\pi^{+}$, and $K_{s}^{0}\pi^{+}\pi^{-}$
modes are required to be within 40 \mev\ of the nominal mass 
\cite{ref:pdg2004}.
In the $K^{-}\pi^{+}\pi^{0}$ decay mode 
the mass is required to be within 70 \mev\ of the nominal mass
\cite{ref:pdg2004}.

The $D^{*0}$ candidates are reconstructed by combining the 
$\Dz$ candidates with a soft $\piz$ or $\gamma$, 
whose momentum in the CM frame is less than 0.45 \gev/c.
The mass difference between $D^{*0}$ and $D^{0}$ ($\Delta M$) is
restricted to be within 0.13--0.17 \gev/$c^2$ and 0.12--0.17 \gev/$c^2$
for $D^{0} \piz$ and $D^{0} \gamma$ modes, respectively. 
We further require that the photon used in $D^{*0} \to \Dz \gamma$ 
reconstruction has a minimum energy of 100 \mev.

Finally $D^{*0} \ell$ candidates are reconstructed by combining 
$D^{*0}$ with an identified electron 
or muon of a momentum above 1.0 \gev/c in the CM frame. 
The $D^{*0}$ and $\ell$ candidates are required to meet at a common vertex.
An additional kinematic constraint is imposed on the reconstructed 
$D^{*0} \ell$ candidates: 
Assuming that the massless neutrino is the only missing particle, we 
calculate the cosine of the angle between the $D^{*0} \ell$ candidate
and the $B$ meson,
\begin{equation}
\cos\theta_{B-D^{*0}\ell} = \frac{2 E_{B} E_{D^{*0}\ell} - m_{B}^{2} - m_{D^{*0}
\ell}^{2}}{2|\vec{p}_{B}||\vec{p}_{D^{*0}\ell}|}.
\end{equation}
Here ($E_{D^{*0}\ell}$, $\vec{p}_{D^{*0}\ell}$) and
($E_{B}$, $\vec{p}_{B}$) are the 
four-momenta in the CM frame, and $m_{D^{*0}\ell}$ and $m_{B}$ 
are the masses of the $D^{*0}\ell$ candidate and $B$ meson, respectively. 
$E_{B}$ and the magnitude of $\vec{p}_{B}$ are calculated 
from the beam energy: $E_{B} = E_{\rm{beam}} = E_{\rm{CM}}/2$ and 
$ | \vec{p}_{B} | = \sqrt{E_{\rm{beam}}^{2} - m_{B}^{2} }$.
Correctly reconstructed candidates
populate the range [-1,1], whereas combinatorial backgrounds
can take unphysical values outside this range. 
We retain events in the interval 
$-1.1 < \cos\theta_{B-D^{*0}\ell} < 1.1$, to take into account
the detector energy and momentum resolution.

If more than one suitable $D^{*0} \ell$ candidate is 
reconstructed in an event, the best candidate is 
selected based on the $\Dz$ mass and $\Delta M$. 
A two dimensional likelihood function is formed by taking
the product of the $\Dz$ mass and $\Delta M$ distributions
obtained from MC simulation. We select the candidate
with the largest likelihood.

The following additional cuts are applied on the selected
best candidate.
The $\Delta M$ value for the best candidate is 
required to be between 0.135--0.150 \gev/$c^2$ and 
0.130--0.155 \gev/$c^2$ for
candidates reconstructed in $\Dz \piz$ and 
$\Dz \gamma$ modes, respectively.
The angle between the $\Dz$ and the 
soft $\piz$ or $\gamma$ from $D^{*0}$ decay in the
CM frame is restricted to be less than 
$60^{\circ}$ and $90^{\circ}$ for $D^{*0}$
reconstructed in $\Dz \piz$ and $\Dz \gamma$ modes, 
respectively. 
The sum of the charge of all the particles in the event (net charge) 
must be equal to zero for events with selected candidates.

\begin{figure}[htb]
     \centering
     \subfigure[The $\Delta M$ distribution of 
               $D^{*0} (\Dz \piz) e \nu$ candidate plotted for 
               \btn \xspace signal simulation (top) and for data and 
               background simulation (bottom).]{
          \label{fig:dmdstarl-d0pi0}
          {\includegraphics[width=.45\textwidth]{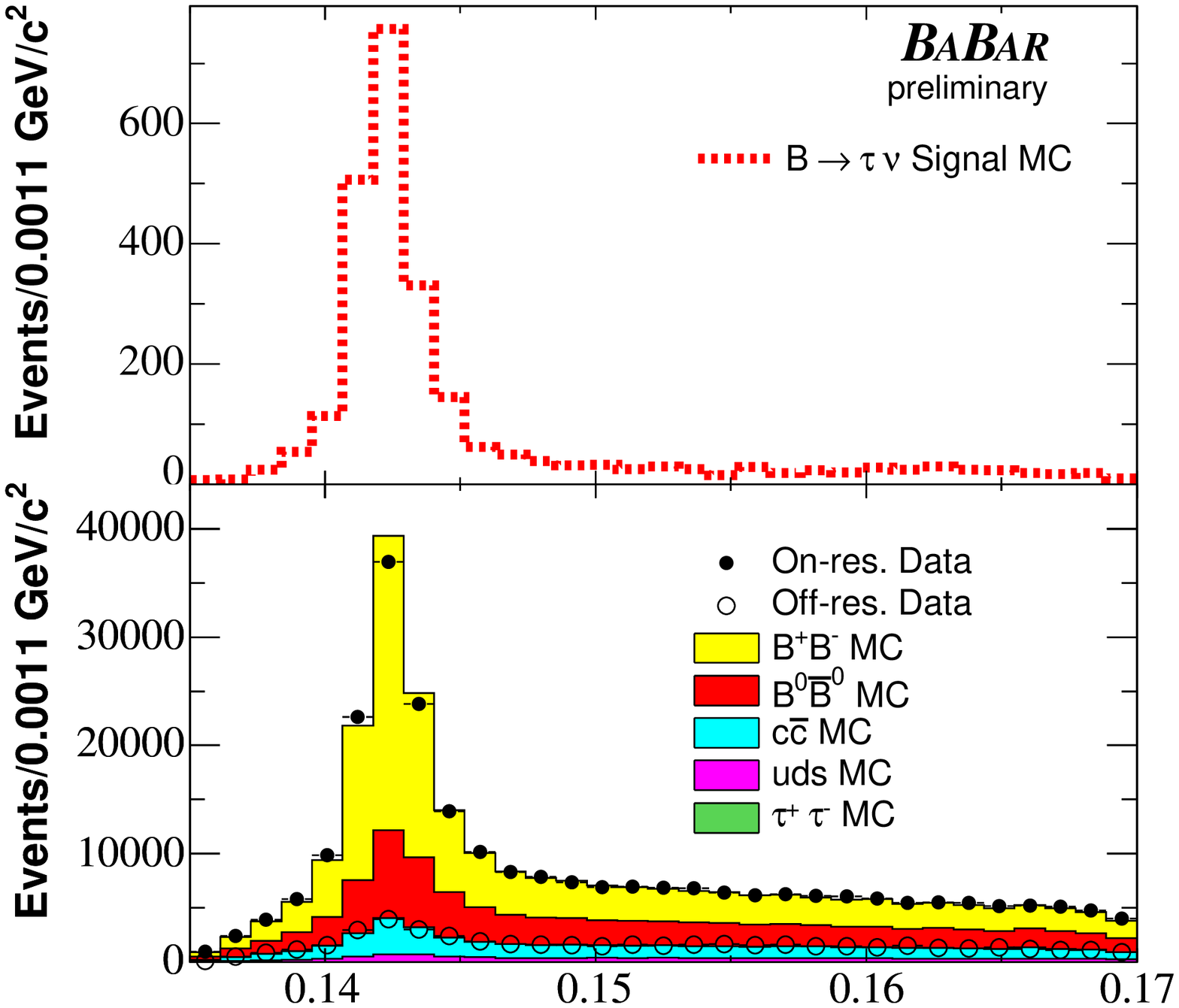}}}
     \hspace{.1in}
     \subfigure[The $\Delta M$ distribution of 
               $D^{*0} (\Dz \gamma) e \nu$ candidate 
               plotted for \btn \xspace signal simulation (top)
               and for data and background simulation (bottom).]{
          \label{fig:dmdstarl-d0gamma}
          {\includegraphics[width=.45\textwidth]{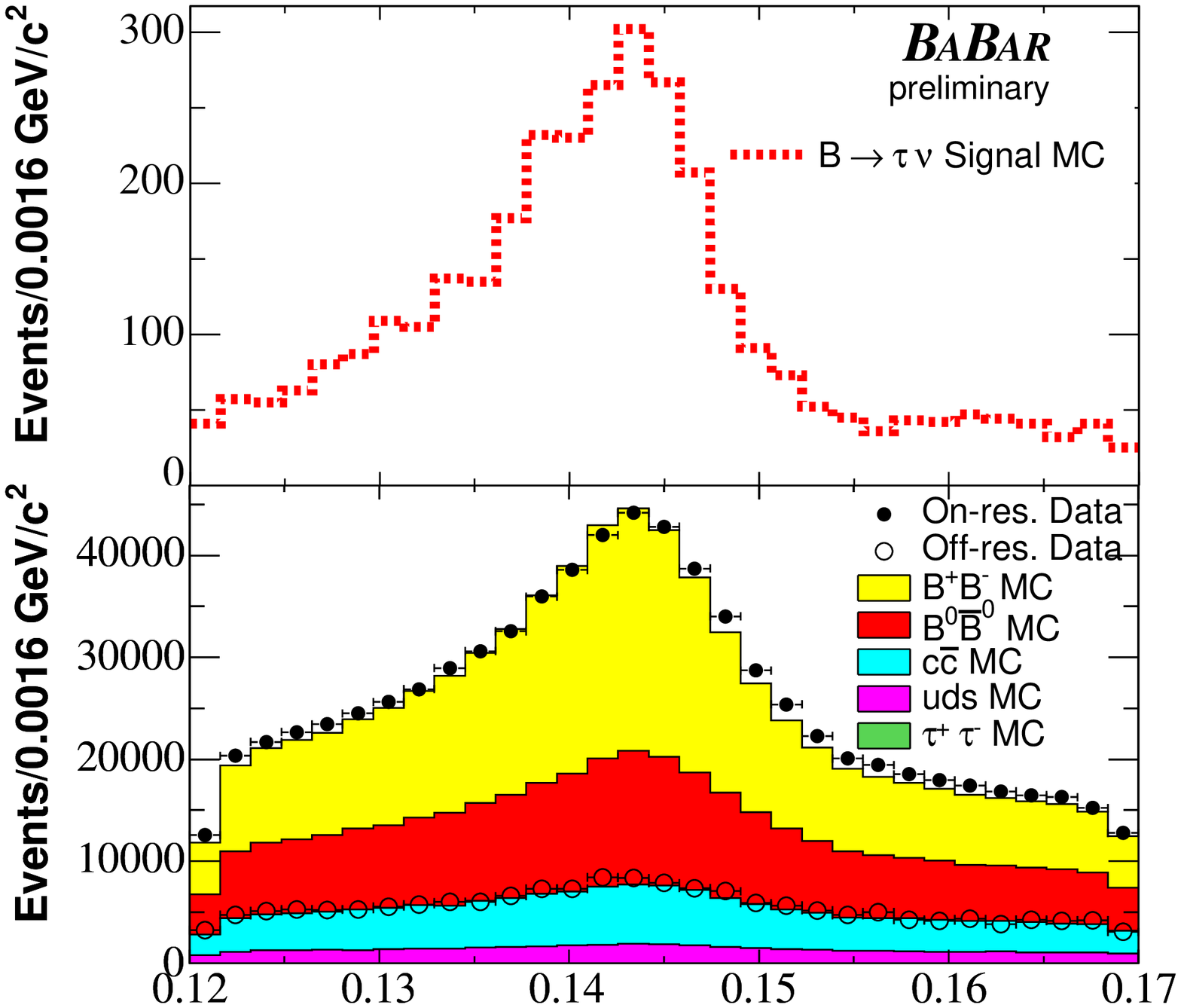}}}\\
     \caption{The $\Delta M$ distributions of $D^{*0} e \nu$ candidate with
              $D^{*0}$ reconstructed in $\Dz \piz$ (fig. \ref{fig:dmdstarl-d0pi0}) and $\Dz \gamma$ 
              (fig. \ref{fig:dmdstarl-d0gamma}) modes,
	      plotted for \btn \xspace simulation, background simulation and data. 
	      The $\Delta M$, and the angle between the $\Dz$ and the soft $\piz$ or $\gamma$ 
              requirements are not applied on the $D^{*0} e \nu$ candidates 
              plotted in these distributions.
Here the simulated background events are scaled to match the data yield. The normalization of the signal MC events is arbitrary. These distributions are similar for the $D^{*0} \mu \nu$ candidates.}
     \label{fig:dmdstarl-Ds0El}
\end{figure}

At this stage of the selection, the observed yield in data and
the predicted yield in the MC simulation agree to 
within approximately 6\%.
This discrepancy is corrected by scaling the yield and efficiency 
obtained from MC simulation.
The scale factor of 0.937 is used to correct 
the tag $B$ reconstruction efficiency in the signal MC simulation.
The systematic error associated with this correction will 
be described in Sec. \ref{sec:Systematics}. The corrected
tag reconstruction efficiency in the signal MC simulation is 
(1.818 $\pm$ 0.074)$\times 10^{-3}$. Figure 
\ref{fig:dmdstarl-Ds0El} shows the $\Delta M$ distributions 
of the selected best $D^{*0} e \nu$ candidates.

\subsection{SELECTION OF \btn \xspace DECAYS}
\label{sec:SigSelection}

\begin{figure}[htb]
     \centering
     \subfigure[Number of signal-side tracks for
               $D^{*0} (\Dz \gamma) e \nu$ tag $B$ candidate plotted for 
               \btn \xspace signal MC simulation (top) and 
               for data and 
               background MC simulation (bottom).]{
          \label{fig:nextrkroedstarlD0gammaBestCandEl}
          {\includegraphics[width=.45\textwidth]{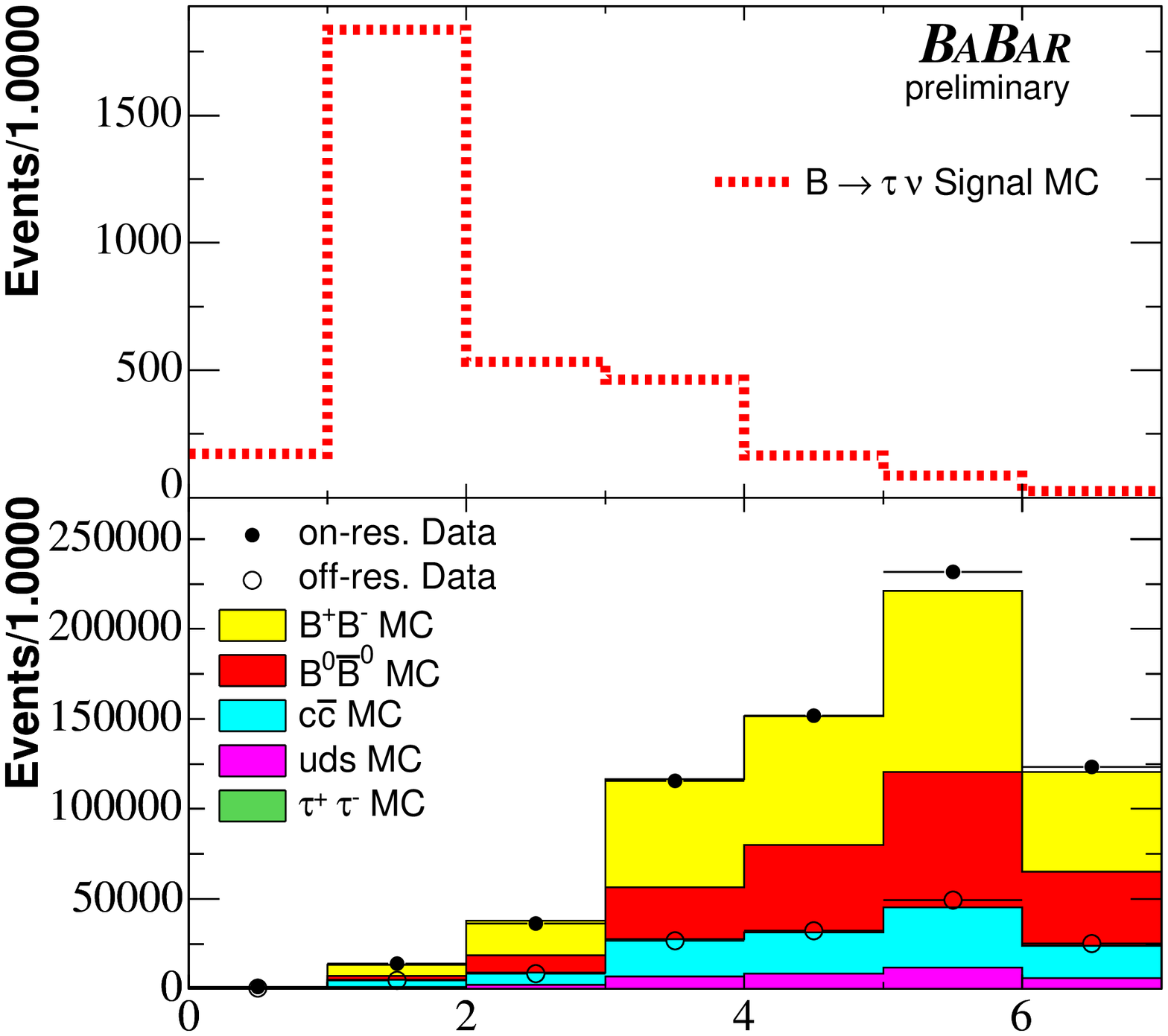}}}
     \hspace{.1in}
     \subfigure[$E_{\rm{extra}}$ distribution after the $D^{*0} (\Dz \gamma) e \nu$ tag $B$ candidate selection 
               plotted for \btn \xspace signal MC simulation (top)
               and for data and background MC simulation (bottom).]{
          \label{fig:Eextra-BestCand}
          {\includegraphics[width=.45\textwidth]{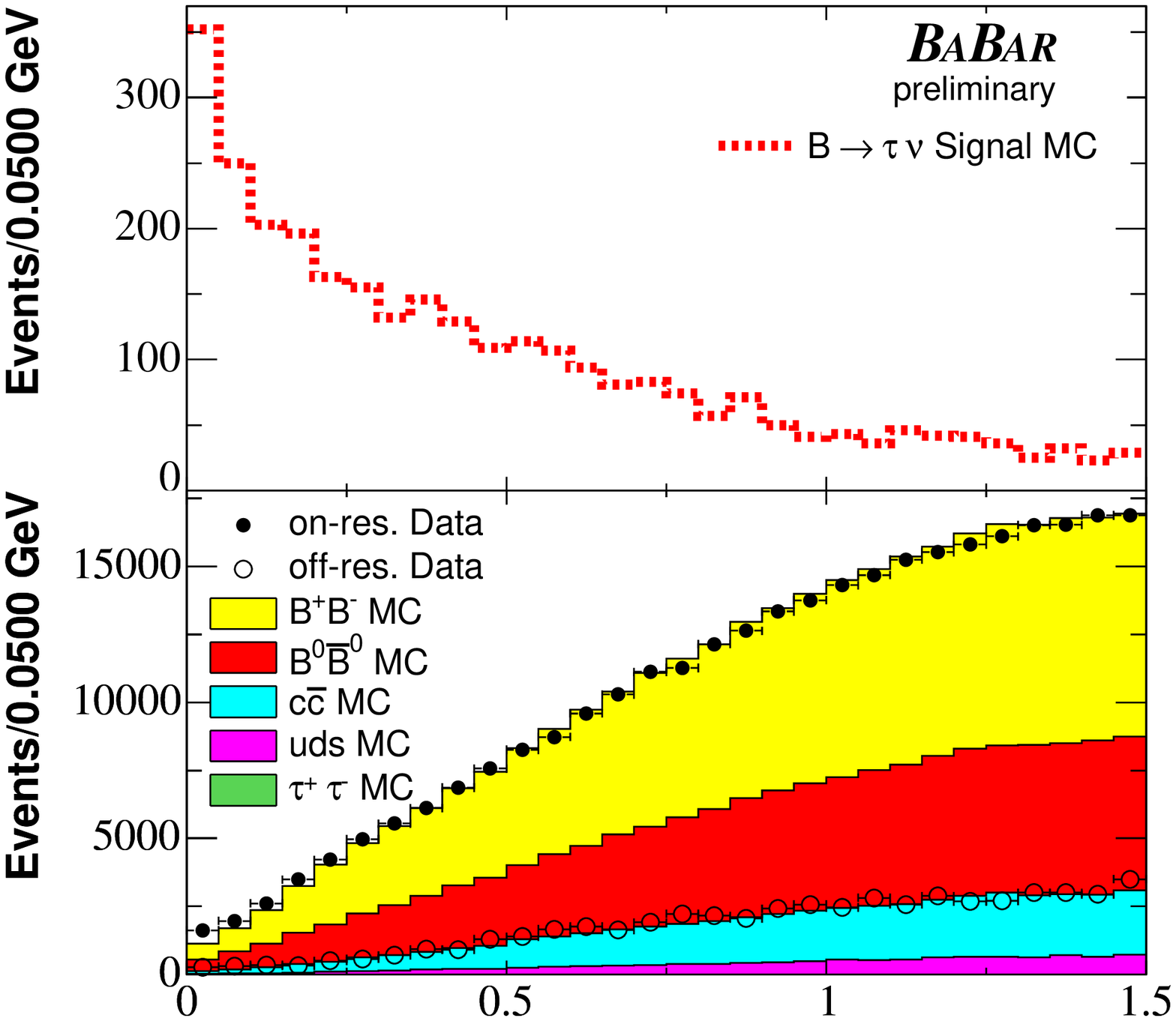}}}\\
     \caption{The number of signal-side tracks and $E_{\rm{extra}}$ distributions
              after tag $B$ selection, plotted for \btn \xspace MC simulation, background MC simulation, and data.
              The net charge requirement is not applied on the events plotted here. 
	      Here the simulated background events are normalized to on-resonance data luminosity. The normalization of the signal MC events is arbitrary. These distributions are similar for other tag $B$ candidates.}
     \label{fig:nSigTrkAndExtra}
\end{figure}

After the tag $B$ reconstruction, in the signal side
the $\tau$ from the $\btn$ decay is identified in one of the following modes:
$\tautoenunu$, $\tautomununu$, $\tautopinu$, $\tautopipiznu$, or
$\tautothreepinu$.
We select events with one or three signal-side track(s).
The event is rejected if any of the signal-side tracks
fail the following selection criteria:
it must have at least 12 DCH hits, its momentum transverse to the 
beam axis, $p_{\rm{T}}$, is greater than 0.1 \gev/c, and
its point of closest approach to the interaction point is 
less than 10.0~\cm\ along the beam axis and less than 1.5~\cm\ transverse 
to the beam axis. Figure \ref{fig:nextrkroedstarlD0gammaBestCandEl}
shows the distribution of the number of signal-side tracks before
this cut. 
The invariant mass of a signal-side $\piz$ candidate
must be between 0.10--0.16 \gev/$c^2$,
the shower shape of the daughter photon candidates must be consistent with 
an electromagnetic shower shape and the photons
must have a minimum energy of 50 \mev.

The most powerful variable for separating signal
and background is the remaining neutral energy ($E_{\rm{extra}}$),
calculated by adding the CM energy of the photons that are not
associated with either the tag $B$ or the $\piz$ candidate from
$\tautopipiznu$ signal decay. The photon candidates contributing
to the $E_{\rm{extra}}$ variable have minimum cluster energies of 20 \mev.
For signal events
the neutral clusters contributing to $E_{\rm{extra}}$ can only
come from processes like  beam-background, hadronic split-offs, and
bremsstrahlung. Therefore the signal events peak at
low $E_{\rm{extra}}$ values and the background events, which contain
additional sources of neutral clusters, are distributed
towards higher $E_{\rm{extra}}$ values (see fig. \ref{fig:Eextra-BestCand}). 
The  $E_{\rm{extra}} < 0.3$ \gev region is defined as the signal region. 
This $E_{\rm{extra}} < 0.35$ \gev region, which is slightly larger than
the signal region, is kept blinded in on-resonance data until the selection 
is optimized.

The different signal modes are distinguished by the following 
selection criteria.
The $\tautoenunu$, $\tautomununu$, $\tautopinu$, and $\tautopipiznu$
signal modes, all of which contain one charged track, are separated by
particle identification.
Both the $\tautopinu$ and the $\tautopipiznu$ modes
contain a pion signal track and are characterized by the 
number of signal-side $\piz$ mesons. The $\tautothreepinu$ mode
contains three signal-side tracks. These signal selection 
requirements are as follows.

\begin{itemize}

\item{Particle identification:}

\begin{itemize} 
\item For the $\tautoenunu$ selection the track must be 
identified as an electron and not identified as a muon or a kaon.

\item For the $\tautomununu$ selection the track must be 
identified as a muon and not identified as an electron or a kaon.

\item For the $\tautopinu$ and $\tautopipiznu$ selection we require that
the track is not identified as an electron or a muon or a kaon.

\item For the $\tautothreepinu$ selection each of the tracks
must be identified as a pion and not identified as
an electron or a muon or a kaon.

\end{itemize} 

\item{Signal-side $\piz$ multiplicity:} 

\begin{itemize}

\item For the $\tautopinu$ selection we require the event to contain 
no signal-side $\piz$. 

\item For the $\tautopipiznu$ selection we require that the 
event contains at least one signal-side $\piz$.

\end{itemize}

\item{$E_{\rm{extra}}$ requirement:}

\begin{itemize}
\item For all the signal modes $E_{\rm{extra}}$ must be less than 0.3 \gev.
\end{itemize}

\end{itemize}

Background consists primarily of $B^{+}B^{-}$ events in which the tag
$B$ meson has been correctly reconstructed. The recoil side contains
one or three track(s) and the additional particles which are not 
reconstructed by the tracking detectors or calorimeters. Typically these
events contain one or more $K_{L}^{0}$ and/or neutrinos, and frequently
also additional charged or neutral particles which pass outside of the 
tracking and calorimeter acceptance. Background events also contain
$B^{0}\overline{B^{0}}$ events. The continuum background contributes to 
hadronic $\tau$ decay modes. In addition some excess events in data,
most likely from two-photon processes which are not modeled in MC 
simulation, are also seen. These backgrounds can be suppressed by the
following constraints on the kinematics of the $\btn$ decays.

\begin{itemize}

\item{Missing mass:} The missing mass is calculated as follows.
\begin{equation}
M_{\rm{miss}} = \sqrt{ (E_{\FourS}-E_{\rm{vis}})^2 - ( \vec{p}_{\FourS} - \vec{p}_{\rm{vis}} )^2 }.
\end{equation}
Here ($E_{\FourS}$, $\vec{p}_{\FourS}$) is the four-momenta of the $\FourS$,
known from the beam energies. The quantities $E_{\rm{vis}}$ and $\vec{p}_{\rm{vis}}$ are the 
total visible energy and momentum of the event which are calculated by adding the 
energy and momentum, respectively, of all the reconstructed 
charged tracks and photons in the event.

\begin{itemize} 

\item For the $\tautoenunu$ and $\tautomununu$ selections, events with 
missing mass less than 4 \gev/$c^2$ are rejected.

\item For the $\tautopinu$ and $\tautopipiznu$ selections, 
the missing mass is required to be greater than 3 \gev/$c^2$.

\item For the $\tautothreepinu$ selection, the missing mass is
required to be greater than 2 \gev/$c^2$.

\end{itemize}

\item{Maximum CM momentum of the $\tau$ daughter:} 

The following maximum CM momentum requirements are applied to 
the $\tau$ daughter particles.

\begin{itemize}

\item The electron candidate from the $\tautoenunu$ decay must have a CM momentum of less than 1.4 \gev/c. 
The CM momentum requirement is not applied to the 
$\tautomununu$ selection because of the following reason:
The momentum spectrum of the lepton from $\tau$ decays peaks below 1 \gev/c.
The particle identification efficiency for low momentum muons is lower than 
that for low momentum electrons. Therefore, applying the maximum momentum 
cut reduces the selection efficiency of the $\tautomununu$ mode 
significantly.

\item For the three hadronic $\tau$ decay modes, 
the $\pi$ from $\tautopinu$, the $\pi \piz$ combination from 
$\tautopipiznu$, or the $3 \pi$ combination from $\tautothreepinu$
must all have CM momenta less than 2.7 \gev/c.

\end{itemize}

\end{itemize}

The $\tautopipiznu$ and $\tautothreepinu$ decays proceed via intermediate 
resonances. For these modes further background rejection can be 
achieved by applying the following requirements on the intermediate mesons.

\begin{itemize}

\item{$\rho^{+}$ selection:}

The $\tautopipiznu$ decay proceeds via an intermediate $\rho^{+}$
state. The signal-side track is combined with a signal-side $\piz$ to form
the $\rho^{+}$ candidate.
In events with more than one signal-side $\piz$, the candidate with 
invariant mass closest to the nominal $\piz$ mass \cite{ref:pdg2004}
is chosen. The invariant mass of the 
reconstructed $\rho^{+}$ is required to be within
0.55--1.00 \gev/$c^2$. A quantity similar to $\cos \theta_{B-D^{*0} \ell}$,
which is defined in section \ref{sec:TagReco}, can be
reconstructed for $\tau \to \rho \nu$ as follows:
\begin{equation}
\cos\theta_{\tau-\rho} = \frac{2 E_{\tau} E_{\rho} - m_{\tau}^{2} - m_{\rho}^{2}}{2|\vec{p}_{\tau}||\vec{p}_{\rho}|},
\end{equation}
where ($E_{\tau}$, $\vec{p}_{\tau}$) and
($E_{\rho}$, $\vec{p}_{\rho}$) are the
four-momenta in the CM frame, $m_{\tau}$ and $m_{\rho}$
are the masses of the $\tau$ and $\rho$ candidate, respectively.
The quantities $|\vec{p}_{\tau}|$ and $E_{\tau}$ are calculated
assuming that the $\tau$ is from the $\btn$ decay, and 
that the $B^{+}$ is almost at rest in the CM frame.
Candidates outside of $-1.1 < \cos\theta_{\tau-\rho} < 1.1$
are excluded.

\item{$\rho^{0}$ and $a_{1}^{+}$ selection:}

The $\tau$ decays to three charged tracks via two intermediate resonances:
$\tau^{+} \to a_{1}^{+} \nutb$,
$a_{1}^{+} \to \rho^{0} \pi^{+}$, and
$\rho^{0} \to \pi^{+} \pi^{-}$. 
The $\pi^{-}\pi^{+}$ combination with an invariant mass closest to the
nominal $\rho^{0}$ mass \cite{ref:pdg2004}
is selected as the best $\rho^{0}$ candidate.
The invariant mass of the selected $\rho^{0}$ must be within
0.55--1.00 \gev/$c^2$. The CM momentum of the
selected $\rho^{0}$ candidate is required to be greater than 0.5 \gev/c.
The invariant mass of the three signal tracks must be within
1.0--1.6 \gev/$c^2$. The total CM momentum of the
three tracks has to be greater than 1.0 \gev/c. The 
three tracks are also required to converge to a common vertex and the 
candidates are rejected if the vertex fit probability is less than 0.1\%.
For $\tau \to a_{1} \nu$ decay the quantity,
\begin{equation}
\cos\theta_{\tau-a_{1}} = \frac{2 E_{\tau} E_{a_{1}} - m_{\tau}^{2} - m_{a_{1}}^{2}}{2|\vec{p}_{\tau}||\vec{p}_{a_{1}}|},
\end{equation}
is obtained using the similar procedure used for calculating
$\cos \theta_{\tau-\rho}$. Here ($E_{\tau}$, $\vec{p}_{\tau}$) and
($E_{a_1}$, $\vec{p}_{a_1}$) are the
four-momenta in the CM frame, $m_{\tau}$ and $m_{a_1}$
are the masses of the $\tau$ and $a_1$ candidate, respectively.
Candidates not satisfying 
$-1.1 < \cos\theta_{\tau-a_{1}} < 1.1$ are excluded.

\end{itemize}


The signal selection criteria for all five signal modes are summarized in Table
\ref{tab:SigSelSummary}. 

\begin{table}[!htb]
\caption{The selection criteria for different signal modes are listed in this table. The symbols $P^{*}_{x}$ and $M_{x}$, used in the table, correspond to the CM momentum and invariant mass of $x$, respectively.}
\vspace{-.3cm}
\begin{center}
    \footnotesize
\begin{tabular}{|c|c|c|c|c|} \hline
$\tautoenunu$ & $\tautomununu$ & $\tautopinu$ & $\tautopipiznu$ & $\tautothreepinu$ \\ \hline \hline
\multicolumn{4}{|c|}{One signal-side track} & Three signal-side tracks \\ \hline
\multicolumn{5}{|c|}{ track quality requirements for each signal track } \\ \hline
electron       & muon         & \multicolumn{2}{|c|}{not electron}    & pion \\ 
not muon       & not electron & \multicolumn{2}{|c|}{not muon}        & not electron \\ 
not kaon       & not kaon     & \multicolumn{2}{|c|}{not kaon}        & not muon  \\ 
               &              & \multicolumn{2}{|c|}{}                & not kaon  \\  \hline
none            &  none         & No signal-side $\pi^{0}$   & Non-zero signal-side $\pi^{0}$  & none          \\  \hline
\multicolumn{5}{|c|}{ $E_{\rm{extra}} < 0.3$ $\gev$} \\ \hline
\multicolumn{2}{|c|}{Missing Mass $>$ 4 $\gev$} & \multicolumn{2}{|c|}{Missing Mass $>$ 3 $\gev$}  &  Missing Mass $>$ 2 $\gev$ \\\hline
$P^{*}$ of      &   none        & $P^{*}$ of        & $P^{*}$ of            & $P^{*}$  of \\
signal-side e   &               & signal-side $\pi$ & signal-side           & of signal-side  \\
track $<$ 1.4 $\gev$   &                    & track $<$ 2.7 $\gev$    & $\pi \pi^{0} < $ 2.7 $\gev$ & 3 $\pi <  2.7$ $\gev$ \\ \hline  
     none        &   none         &  none         &  $\rho^{+}$ selection:               & $a_{1}^{+}$ selection: \\
                 &                &               &  0.55 $< M_{\rho^{+}}< $ 1 $\gev$    & 0.55 $ < M_{\pi^+ \pi^-} <$ 1.0 $\gev$ \\ 
                 &                &               &  $-1.1 < \cos\theta_{\tau-\rho} < 1.1$  & $P^{*}_{\pi^+ \pi^-} >$ 0.5 $\gev$ \\ 
                 &                &               &                                        & 1.0 $ < M_{3 \pi} < $ 1.6 $\gev$ \\ 
                 &                &               &                                        & $P^{*}_{3 \pi} >$ 1.0 $\gev$ \\  
                 &                &               &                                        & Vertex prob. of 3 tracks $> 1 \%$ \\  
                 &                &               &                                        & $-1.1 < \cos\theta_{\tau-a_{1}} < 1.1$ \\ \hline 
\end{tabular}
  \label{tab:SigSelSummary}
\end{center}
\end{table}


\subsubsection{SIGNAL EFFICIENCY}
\label{sec:SigEff}

The ``signal-side selection efficiencies'' 
for the $\tau$ decay modes are 
determined from signal MC simulation and summarized 
in Table \ref{tab:Comb-Final-SigSelEff}.
For each $\tau$ selection mode, the signal-side
efficiency ($\varepsilon_{i}$, where $i$ $\equiv$ selection mode)
is computed as the ratio of the number of events surviving
the requirements of that selection mode
to the number of events where a tag $B$ meson is reconstructed.
In the computation of the total 
signal-side efficiency for each selection 
we take into account the cross-feed from other $\tau$ decay modes reported in 
Table \ref{tab:Comb-Final-SigSelEff}.

The selection efficiency for $\tautomununu$ is low compared to that of the
$\tautoenunu$ mode, because of the fact that the momentum spectrum 
of the signal muons peaks below 1 \gev/c, where the muon detection
efficiency is low. Since no minimum momentum requirement and no tight pion
identification criteria are applied to the
$\tautopinu$ signal selection, electron and muon signal tracks 
that fail particle identification requirement get selected in this mode. 
Any true $\tautopipiznu$ signal events, with a missed $\piz$ 
also get included in $\tautopinu$ selection mode. 
Therefore the $\tautopinu$ selection
mode has the highest signal efficiency. 
From MC estimation we expect $\sim$10 signal events at 112.5 $\rm{fb^{-1}}$,
assuming $\mathcal{B}(\btn)=10^{-4}$.

\begin{table}[!htb]
\caption{Efficiency of the different selections (columns) for the most abundant $\tau$ decay modes (rows). The last two rows show the total efficiency for each selection weighted by the decay branching fractions, and the total efficiency. The errors are statistical only. The total efficiency for each selection is $\varepsilon_{i} = \Sigma_{j=1}^{8} \varepsilon_{i}^{j} f_{j}$, where $\varepsilon_{i}^{j}$ is the efficiency of the selection $i$ for the simulated $\tau$ decay mode $j$. The index $j$ corresponds to the different $\tau$ decay mode in the MC simulation, and $f_{j} = \mathcal{B}(\tau \to j)$ are the $\tau$ branching fractions from Ref. \cite{ref:pdg2004}. If the efficiency is zero a 90\% upper limit is quoted.}
\begin{center}
\begin{tabular}{|p{1.0in}|c|c|c|c|c|} \hline 
$\tau$ decay & $ e \nu \nu $ & $\mu\nu\nu$  & $\pi\nu$ & $\pi\pi^{0}\nu$ & $\pi\pi\pi\nu$ \\  
mode in      & selection     & selection    & selection   & selection    & selection \\ 
Monte Carlo  & efficiency    & efficiency   & efficiency  & efficiency   & efficiency  \\
simulation   &  (\%)         &  (\%)        &  (\%)       &  (\%)        &  (\%) \\  \hline \hline 
$e \nu \overline{\nu}$   & 48.6 $\pm$ 2.8 & 0.1 $\pm$ 0.1  & 11.9 $\pm$ 1.2   & 0.1 $\pm$ 0.1   & 0 ($<0.2$)  \\ \hline 
$\mu \nu \overline{\nu}$ & 0.1 $\pm$ 0.1  & 25.8 $\pm$ 1.9 & 53.3 $\pm$ 3.0   & 0.7 $\pm$ 0.3   & 0 ($<0.3$)  \\ \hline 
$\pi \nu$                & 0 ($<0.4$)     & 0.5 $\pm$ 0.3  & 57.5 $\pm$ 4.0   & 2.6 $\pm$ 0.7   & 0 ($<0.4$)  \\ \hline 
$\pi \pi^{0} \nu$        & 0 ($<0.2$)     & 0.3 $\pm$ 0.1  & 12.1 $\pm$ 1.0   & 8.8 $\pm$ 0.8   & 0 ($<0.2$)  \\ \hline 
$\pi \pi \pi \nu$        & 0 ($<0.5$)     & 0 ($<0.5$)     & 1.1 $\pm$ 0.6    & 0 ($<0.5$)      & 27.6 $\pm$ 3.2  \\ \hline 
$\pi \pi^{0} \pi^{0}  \nu$ & 0 ($<0.5$)   & 0 ($<0.5$)     & 2.8 $\pm$ 0.8    & 2.8 $\pm$ 0.8   & 0 ($<0.5$)  \\ \hline 
$\pi \pi \pi \pi^{0} \nu$ & 0 ($<1.0$)    & 0 ($<1.0$)     & 0.5 $\pm$ 0.5    & 0 ($<1.0$)      & 4.1 $\pm$ 1.5  \\ \hline 
other & 1.2 $\pm$ 0.5 & 2.1 $\pm$ 0.7 & 10.0 $\pm$ 1.5   & 2.5 $\pm$ 0.7   & 1.0 $\pm$ 0.4  \\ \hline \hline 
all $\tau$ decay & 8.4 $\pm$ 0.4 & 4.8 $\pm$ 0.3 & 21.9 $\pm$ 0.7   & 3.2 $\pm$ 0.2   & 2.1 $\pm$ 0.2  \\ \hline 
 Total            & \multicolumn{5}{|c|}{  40.4 $\pm$ 0.9 } \\ \hline 
\hline  
\end{tabular}

\end{center}
\label{tab:Comb-Final-SigSelEff}
\end{table}

\subsubsection{EXPECTED BACKGROUND FROM MONTE CARLO SIMULATION}
\label{sec:MCBkgEst}

To obtain the background estimation from the MC simulation, \BB\ and 
$\epem \to$ \uubar, \ddbar, \ssbar, \ccbar, and \tautau\ events are
scaled to equivalent luminosity in data. The estimated background 
in different selection modes is listed in Table   
\ref{tab:SummaryMCBkgEstSigYield}.
The three modes $\tautoenunu$, $\tautomununu$, $\tautopinu$ are 
relatively clean compared to the other two selection
modes. Signal to background ratios in $\tautoenunu$,
$\tautomununu$, and $\tautopinu$ selection modes are 
10:59, 10:75, and 10:85 respectively, 
assuming $\mathcal{B}(\btn) = 10^{-4}$.
For the $\tautopipiznu$ and $\tautothreepinu$ modes the
signal to background ratio is 10:354 and 10:513, respectively.

\begin{table}[!htb]
\caption{Expected final raw background and signal yield at \onlumi \xspace estimated from background and signal MC simulation. No systematic correction is applied on simulated events. The listed errors are statistical only. If the efficiency is zero a 90\% upper limit on the expected background is quoted.}
\vspace{-.30cm}
\begin{center}
\begin{tabular}{|p{1.2in}|c|c|c|c|c|} \hline 
    &  $\enunu$ &  $\mununu$ & $\pinu$ & $\pipiznu$ & $\threepinu$ \\  \hline \hline 
$\Bu\Bub$  & 8.29 $\pm$ 1.69  & 7.25 $\pm$ 1.58  & 24.18 $\pm$ 2.89  & 14.16 $\pm$ 2.21  & 14.16 $\pm$ 2.21  \\ \hline 
$\BzBzb$  & 3.36 $\pm$ 0.97  & 1.40 $\pm$ 0.63  & 8.11 $\pm$ 1.51  & 3.64 $\pm$ 1.01  & 7.83 $\pm$ 1.48  \\ \hline 
Combined   & &  &  &  &              \\ 
$B \overline{B}$ & 11.65 $\pm$ 1.95 & 8.65 $\pm$ 1.70 & 32.29 $\pm$ 3.26 & 17.80 $\pm$ 2.43 & 21.99 $\pm$ 2.66 \\ \hline 
\hline 
$c \overline{c}$  & 0 ($<2.2$)  & 0 ($<2.2$)  & 9.54 $\pm$ 3.02  & 8.59 $\pm$ 2.86  & 2.86 $\pm$ 1.65  \\ \hline 
$u \overline{u}$, $d \overline{d}$, $s \overline{s}$ & 0 ($<1.9$)  & 0 ($<1.9$)  & 0 ($<1.9$)  & 0 ($<1.9$)  & 0.81 $\pm$ 0.81  \\ \hline 
$\tau^{+} \tau^{-}$  & 0.48 $\pm$ 0.48  & 0 ($<1.1$)  & 2.87 $\pm$ 1.17  & 0.48 $\pm$ 0.48  & 0 ($<1.1$)  \\ \hline 
Combined & &  &  &  &              \\  
non $B \overline{B}$ & 0.48 $\pm$ 0.48 & 0            & 12.41 $\pm$ 3.24 & 9.07 $\pm$ 2.90 & 3.67 $\pm$ 1.84 \\ \hline 
\hline 
Total Background     & 12.12 $\pm$ 2.01 & 8.65 $\pm$ 1.70 & 44.69 $\pm$ 4.59 & 26.86 $\pm$ 3.79 & 25.67 $\pm$ 3.24 \\ \hline 
\hline 
Expected signal      & &  &  &  &            \\   
events for           & 2.03 $\pm$ 0.10  & 1.16 $\pm$ 0.07  & 5.26 $\pm$ 0.16  & 0.76 $\pm$ 0.06  & 0.50 $\pm$ 0.05  \\  
\BR($\btn$) = $10^{-4}$ & &  &  &  &           \\   \hline 
\hline  
\end{tabular} 

  \label{tab:SummaryMCBkgEstSigYield}
\end{center}
\end{table}

\subsubsection{BACKGROUND ESTIMATION FROM $E_{\rm{extra}}$ SIDE BAND IN DATA}
\label{sec:EextraSBExtrapolation}

The $E_{\rm{extra}} < 0.3$ \gev\ region is defined as the ``signal
region'' and the $0.35 < E_{\rm{extra}} < 1.0$ \gev\ region
is defined as the ``side band''. 
The $E_{\rm{extra}}$ shape in the MC distribution is used to extrapolate the data 
side band to the signal region.

The number of MC events in signal region ($N_{\rm{Sig}}^{\rm{MC}}$)
and side band ($N_{\rm{SideB}}^{\rm{MC}}$) are counted
and their ratio ($R_{\rm{MC}}$) is obtained.

\begin{eqnarray*}
R_{\rm{MC}} & = & \frac{N_{\rm{Sig}}^{\rm{MC}}}{N_{\rm{SideB}}^{\rm{MC}}}
\end{eqnarray*}

\noindent Using the number of data events in the side band ($N_{\rm{SideB}}^{\rm{data}}$)
and the ratio $R_{\rm{MC}}$, the number of expected background events in the
signal region in data ($N_{\rm{SideB}}^{\rm{data}}$) is estimated.

\begin{eqnarray*}
N_{\rm{SideB}}^{\rm{data}} & = & N_{\rm{SideB}}^{\rm{data}} \cdot R_{\rm{MC}}
\end{eqnarray*}

\noindent The background estimation for the different selection modes from 
the $E_{\rm{extra}}$ side band extrapolation are shown in Table 
\ref{tab:BkgEst_EleftSB_Extrapolation}. The number of estimated 
background events in the signal region from the data side band extrapolation 
are in agreement with the background estimation from MC simulation within 
statistical uncertainty.

\begin{table}[!htb]
\caption{Background estimation in the signal region  
($E_{\rm{extra}} <$ 0.3 \gev) for the different selection modes.}
\vspace{-1.0cm}
\begin{center}
\begin{tabular}{|p{0.7in}|c|c|c|c|c|} \hline

         &  &  &  & &   \\ 
 Selection    & $N^{\rm{MC}}_{\rm{SideB}}$ & $N^{\rm{MC}}_{\rm{Sig}}$ & $R_{\rm{MC}}$ & $N^{\rm{data}}_{\rm{SideB}}$  & $N^{\rm{exp}}_{\rm{Sig}}$  \\
         &  &  &  & &   \\ \hline \hline

$\enunu$ & 83.63 $\pm$  5.67 & 12.12 $\pm$  2.01 & 0.14 $\pm$  0.03 & 103.00 $\pm$  10.15 & 14.93 $\pm$  3.05     \\ \hline
$\mununu$ & 64.18 $\pm$  4.83 & 8.65 $\pm$  1.70 & 0.13 $\pm$  0.03 & 53.00 $\pm$  7.28 & 7.14 $\pm$  1.80        \\ \hline
$\pinu$ & 227.09 $\pm$  10.86 & 44.69 $\pm$  4.59 & 0.20 $\pm$  0.02 & 250.00 $\pm$  15.81 & 49.20 $\pm$  6.39    \\ \hline
$\pipiznu$ & 179.44 $\pm$  9.37 & 26.86 $\pm$  3.79 & 0.15 $\pm$  0.02 & 182.00 $\pm$  13.49 & 27.24 $\pm$  4.57  \\ \hline
$\threepinu$ & 218.59 $\pm$  9.55 & 25.67 $\pm$  3.24 & 0.12 $\pm$  0.02 & 199.00 $\pm$ 14.11 & 23.37 $\pm$ 3.53  \\ \hline
\hline
\end{tabular}

\label{tab:BkgEst_EleftSB_Extrapolation}
\end{center}
\end{table}

\section{VALIDATION OF $E_{\rm{extra}}$ SIMULATION}
\label{sec:EextraValidation}

The $E_{\rm{extra}}$ distribution in signal and background MC simulation 
are validated using various control samples.
We compare the $E_{\rm{extra}}$ distributions between on-resonance data and 
MC simulation, in both signal region and side band, using 
the control samples. Agreement between the distributions would provide 
validation of the $E_{\rm{extra}}$ modeling in the simulation.

\subsection{$E_{\rm{extra}}$ IN THE SIGNAL MONTE CARLO SIMULATION}
\label{sec:SigMCEextraValidation}

The ``double-tagged'' events, for which
both of the $B$ mesons are reconstructed in tagging modes, 
$B^{-} \rightarrow D^{*0} \ell^{-} \overline{\nu}_{\ell}$ vs 
$B^{+} \rightarrow \overline{D}^{*0} \ell^{+} \nu_{\ell}$, 
are used as a control sample to validate the $E_{\rm{extra}}$
simulation. Due to the large branching fraction and
high tagging efficiency for these events, a sizable 
sample of such events 
is reconstructed in the on-resonance dataset. 
These double-tag events contain very little background due to the
full reconstruction of the event. 

To select double-tag events 
we require that the two tag $B$ candidates do not share 
any tracks or neutrals. If there are more than two such 
non-overlapping tag $B$ candidates in the event then the 
best two are selected using
the same best candidate selection criteria, as described in
  Sec. \ref{sec:TagReco}.
After selecting the two tag $B$ candidates only
the events with no extra charged tracks are selected.

\begin{figure}[htp]
     \centering
     {\includegraphics[width=.4\textwidth]{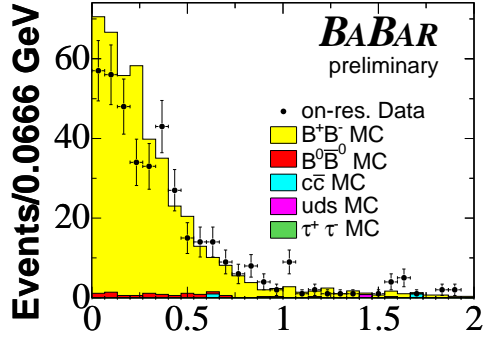}}
     \caption{The distribution of the remaining neutral energy ($E_{\rm{extra}}$) for double-tagged events, plotted for simulation and data.}
     \label{fig:DoubleTageCNSum}
\end{figure}

The $E_{\rm{extra}}$ for the double-tagged sample is calculated by summing the 
CM energy of the photons which are not associated with 
either of the tag $B$ candidates. The sources of neutrals contributing
to the $E_{\rm{extra}}$ distribution in double-tagged events  
are similar to those contributing to the $E_{\rm{extra}}$ distribution  
in the signal MC simulation. Therefore the agreement of
the $E_{\rm{extra}}$ distribution between data and MC simulation for the 
double-tagged sample, in figure \ref{fig:DoubleTageCNSum},
is used as a validation of the $E_{\rm{extra}}$ simulation in the signal 
MC.  

\subsection{$E_{\rm{extra}}$ IN THE BACKGROUND MONTE CARLO SIMULATION}
\label{sec:BkgMCEextraValidation}

The following two background control samples are used to study the 
agreement between data and simulation in the $E_{\rm{extra}} < 0.35$ \gev\ 
region. 

\begin{itemize}

\item Events with two remaining signal-side tracks

\item Events with non-zero net charge

\end{itemize}

The procedure from Sec. \ref{sec:EextraSBExtrapolation},
the background estimation from the $E_{\rm{extra}}$ side band, is used 
for this test. In Table \ref{tab:SideBand_Extrapolation} 
we show the comparison of the number of expected data 
events ($N_{\rm{SideB}}^{\rm{data}}$) in the signal region with the observed number of data
events ($N_{\rm{Sig}}^{\rm{obs}}$) in the signal region. 
The agreement between
the above two quantities provides validation of background estimation in the
low $E_{\rm{extra}}$ region.

\begin{table}[!htb]
\caption{Test of background estimation in the low $E_{\rm{extra}}$ region from various control samples. Signal selection cuts for different $\tau$ decay modes are applied on the control samples. The expected and observed number of events in the signal region agree within error for all the samples ($P(\chi^2)=0.15$ between the entries in the last two columns. Here $\chi^2 = 10.77$, and the number of degrees of freedom is 7.).}
\vspace{-1.0cm}
\begin{center}
\begin{tabular}{|p{0.6in}|c|c|c|c|c|} \hline 
                &  &  &  & &   \\  Selection    & $N_{\rm{SideB}}^{\rm{MC}}$ & $N_{\rm{SideB}}^{\rm{data}}$  & $N_{\rm{Sig}}^{\rm{MC}}$ & $N_{\rm{Sig}}^{\rm{exp}}$ & $N_{\rm{Sig}}^{\rm{obs}}$ \\                 &  &  &  & &   \\ \hline \hline 

\multicolumn{6}{|c|}{} \\
\multicolumn{6}{|c|}{Two signal-side track control sample} \\
\hline

$e \nu \overline{\nu}$ & 223.34 $\pm$ 8.68 & 203.00 $\pm$ 14.25 & 76.91 $\pm$ 5.12 & 69.90 $\pm$ 7.29 & 64.00 $\pm$ 8.00 \\ \hline 
$\mu \nu \overline{\nu}$ & 144.80 $\pm$ 7.14 & 123.00 $\pm$ 11.09 & 57.68 $\pm$ 4.43 & 49.00 $\pm$ 6.29 & 55.00 $\pm$ 7.42 \\ \hline 
$\pi \nu $ & 1132.39 $\pm$ 21.64 & 1050.00 $\pm$ 32.40 & 401.89 $\pm$ 12.53 & 372.65 $\pm$ 17.83 & 422.00 $\pm$ 20.54 \\ \hline 
$\pi \pi^{0} \nu $ & 1175.99 $\pm$ 21.78 & 959.00 $\pm$ 30.97 & 241.36 $\pm$ 9.70 & 196.83 $\pm$ 10.78 & 232.00 $\pm$ 15.23 \\ \hline 

\multicolumn{6}{|c|}{} \\
\multicolumn{6}{|c|}{$Q_{net} \neq 0$ control sample} \\
\hline

$e \nu \overline{\nu}$ & 16.86 $\pm$ 2.33 & 8.00 $\pm$ 2.83 & 4.10 $\pm$ 1.14 & 1.94 $\pm$ 0.92 & 5.00 $\pm$ 2.24 \\ \hline 
$\mu \nu \overline{\nu}$ & 6.66 $\pm$ 1.46 & 2.00 $\pm$ 1.41 & 2.29 $\pm$ 0.87 & 0.69 $\pm$ 0.57 &    0           \\ \hline 
$\pi \nu$ & 68.62 $\pm$ 5.31 & 63.00 $\pm$ 7.94 & 23.26 $\pm$ 2.83 & 21.36 $\pm$ 4.09 & 24.00 $\pm$ 4.90 \\ \hline 
$\pi \pi^{0} \nu$ & 63.14 $\pm$ 5.14 & 44.00 $\pm$ 6.63 & 9.14 $\pm$ 1.74 & 6.37 $\pm$ 1.63 & 6.00 $\pm$ 2.45 \\ \hline 
\hline  
\end{tabular}

\label{tab:SideBand_Extrapolation}
\end{center}
\end{table}

\section{STUDIES OF SYSTEMATICS}
\label{sec:Systematics}

The main sources of uncertainty in the determination of the $\btn$
branching fraction are the following:

\begin{itemize}

\item We divide the uncertainties in signal efficiency determination into two categories.
\begin{itemize}
\item Uncertainty in tagging efficiency determination 

\item Uncertainty in determination of the efficiency $\varepsilon_{i}$
for each selection mode.
\end{itemize}

\item Uncertainty in the determination of the number of expected background
events in the signal region for each selection mode.

\end{itemize}

A small uncertainty of 1.1\% also enters the branching ratio limit calculation 
from the estimation of the number of $B^{+}B^{-}$ events present in the 
data sample \cite{bib:BCount}.

\subsection{TAGGING EFFICIENCY SYSTEMATICS}
\label{sec:tagEffSys}

The tagging efficiency and yield in signal simulation is corrected 
using the ``double-tagged'' events. The selection of 
``double-tagged'' events is described
in sec. \ref{sec:SigMCEextraValidation}.

The number of double-tagged events ($N_{2}$) is given by
\begin{eqnarray}
\label{eqn:dbltageff}%
N_2 & = & \varepsilon^{2} N
\end{eqnarray}
where $\varepsilon$ is the efficiency and $N$ is the original number
of events. 
The double tag yield in data is 407.0 $\pm$ 20.2.
In the simulation, we find 434.4 $\pm$ 12.4 double-tags, 
with very few non-$\Bu\Bub$\ events. 
The expected number of $\Bu\Bub$ events in the dataset is
$N = (62.06 \pm 0.68)\times10^{6}$ events.
Calculating the efficiency using the numbers of double-tags in 
data and normalized MC in
equation \ref{eqn:dbltageff}, we find the efficiencies $\varepsilon_{\rm{data}}$
and $\varepsilon_{\rm{MC}}$. The correction factor, ratio of the efficiencies between data and simulation, from this method is
\begin{eqnarray*}
\frac{\varepsilon_{\rm{data}}}{\varepsilon_{\rm{MC}}} & = & 0.969 \pm 0.029.
\end{eqnarray*}

The correction factor obtained from the double tags
agrees within error with the normalization
factor used in section \ref{sec:TagReco} to correct tag $B$ meson
yield in simulation. Therefore, the double tags provide the 
validation for the tag $B$ yield correction. We take the 3.1\% 
error obtained from the double tag method as the systematic error
associated with the tagging efficiency and yield correction in Monte
Carlo.

\subsection{UNCERTAINTIES IN THE SIGNAL SELECTION EFFICIENCIES IN EACH SELECTION MODE}
\label{sec:sigEffSys}

Besides tagging efficiency uncertainty,
the contribution to the systematic uncertainties in the determination of the 
efficiencies comes from systematic uncertainty on the tracking efficiency,
particle identification, and simulation of the neutral clusters in the 
calorimeter which contribute to the $E_{\rm{extra}}$ distribution. 
Uncertainty in the $\piz$ reconstruction efficiency introduces
an additional contribution to the systematics in the $\tautopinu$ (5.0\%)
and $\tautopipiznu$ (7.7 \%) selection modes. The different contributions 
to the systematic uncertainty on the selection efficiencies are
listed in table \ref{tab:SignalEffSys}.

\begin{table}[!htb]
\caption{Contribution to the systematic uncertainty on the signal selection efficiencies in different selection modes.}
\begin{center}
\begin{tabular}{|c|c|c|c|c|c|} \hline
Selection     &  tracking & Particle       & Neutral            & Total       &  Correction \\
modes         &   (\%)    & Identification & Reconstruction     & Systematic  &  Factor      \\
              &           &      (\%)      &     (\%)           & Error (\%)  &              \\
\hline \hline
$\enunu$      &  1.4      &  0.1           &  3.1               &     3.4     &  0.99   \\ \hline
$\mununu$     &  1.4      &  3.0           &  2.3               &     4.0     &  0.89    \\ \hline
$\pinu$       &  1.4      &  0.5           &  3.1 $\oplus$ 5.0  &     6.1     &  1.02    \\ \hline
$\pipiznu$    &  1.4      &  0.1           &  2.9 $\oplus$ 7.7  &     8.3     &  0.95    \\ \hline
$\threepinu$  &  4.2      &  0.3           &  4.7               &     6.3     &  1.00    \\ \hline
\end{tabular}
\end{center}
\label{tab:SignalEffSys}
\end{table}

\subsection{UNCERTAINTIES IN THE BACKGROUND ESTIMATION}
\label{sec:bkgEstSys}

The background estimation is performed by extrapolating the 
number of events in the $E_{\rm{extra}}$ side band in data into the 
signal region as described in sec. \ref{sec:EextraSBExtrapolation}.
The major uncertainty related to background 
estimation comes from the data and MC statistics.
The modeling of the $E_{\rm{extra}}$ variable in the background
MC contributes to additional systematic corrections
to the background estimation. The systematic corrections 
due to $E_{\rm{extra}}$ modeling for
different modes are the following:  
(1.02 $\pm$  0.04) for $\tautoenunu$, 
(1.13 $\pm$  0.06) for $\tautomununu$, 
(1.12 $\pm$  0.03) for $\tautopinu$, 
(1.09 $\pm$  0.04) for $\tautopipiznu$ and 
(1.07 $\pm$  0.03) for $\tautothreepinu$.

\section{RESULTS}
\label{sec:Physics}

\begin{figure}[htp]
     \centering
     \subfigure[The distribution of the $E_{\rm{extra}}$ 
     	for events satisfying $\tautoenunu$ selection.]{
          \label{fig:EextraSB_enunu_eCNSum}
          {\includegraphics[width=.31\textwidth]{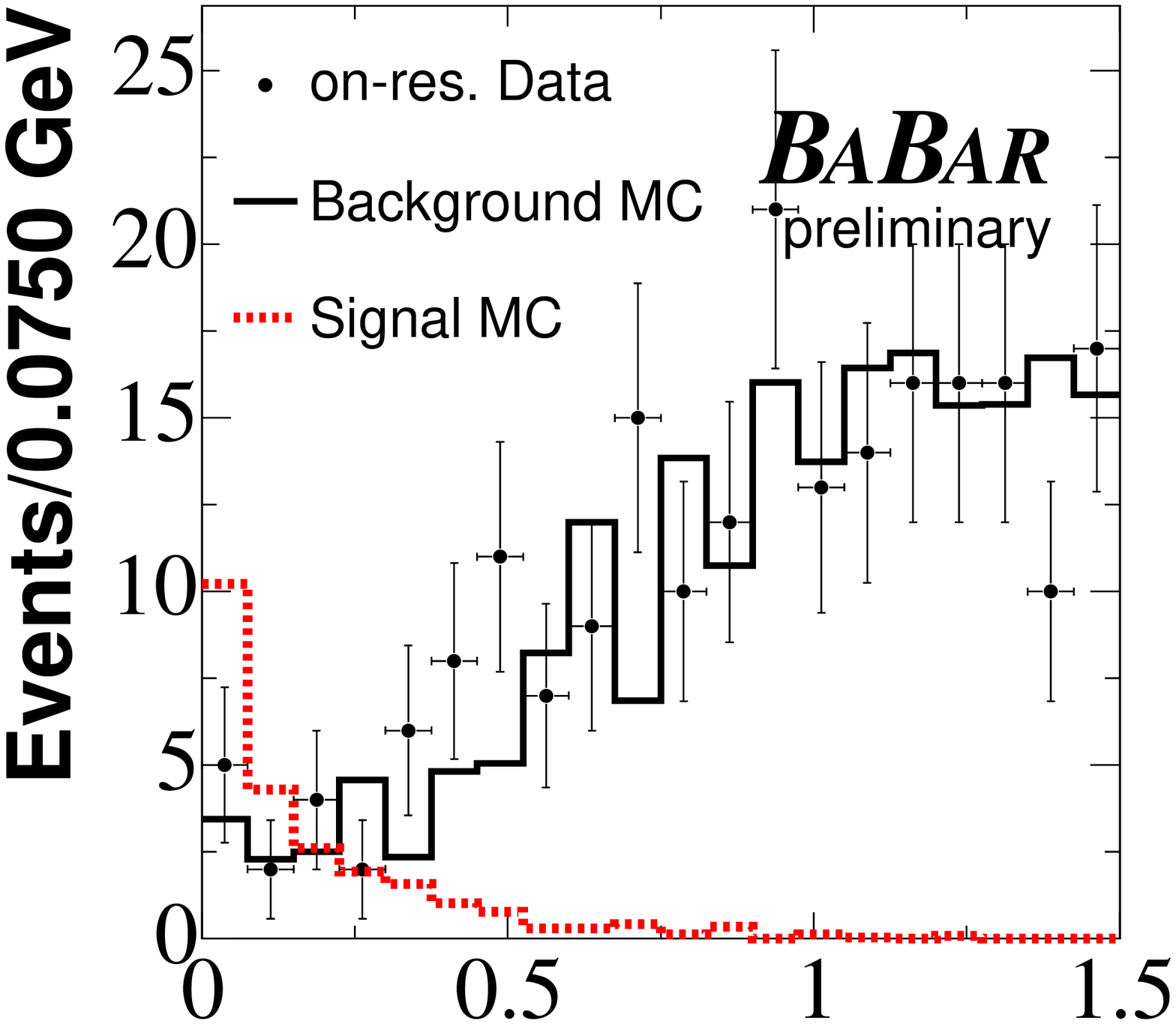}}}
     \hspace{.02in}
      \subfigure[The distribution of the $E_{\rm{extra}}$ 
     	for events satisfying $\tautomununu$ selection.]{
          \label{fig:EextraSB_mununu_eCNSum}
          {\includegraphics[width=.31\textwidth]{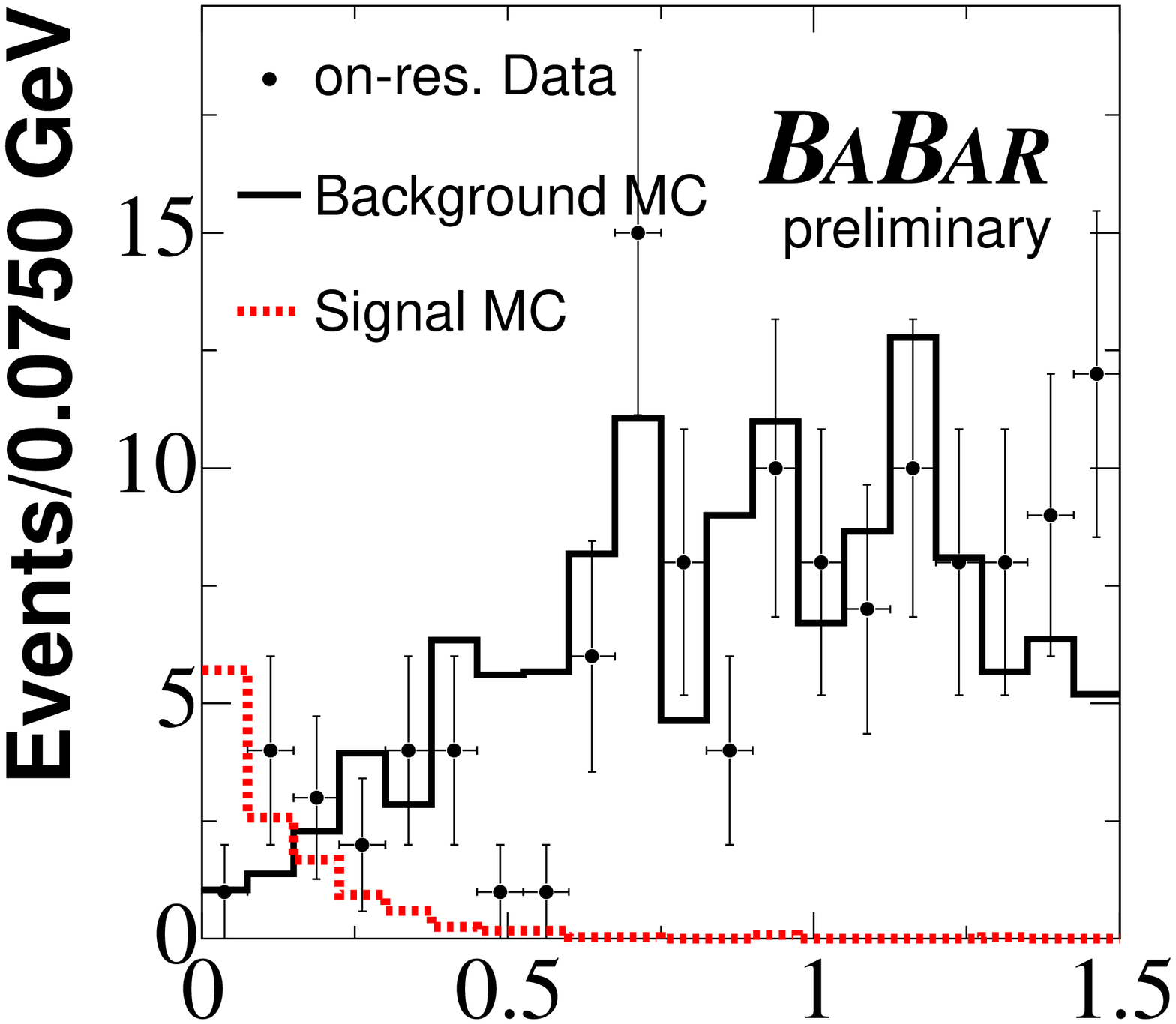}}}
     \hspace{.02in}
      \subfigure[The distribution of the $E_{\rm{extra}}$ 
     	for events satisfying $\tautopinu$ selection.]{
          \label{fig:EextraSB_pinu_eCNSum}
          {\includegraphics[width=.31\textwidth]{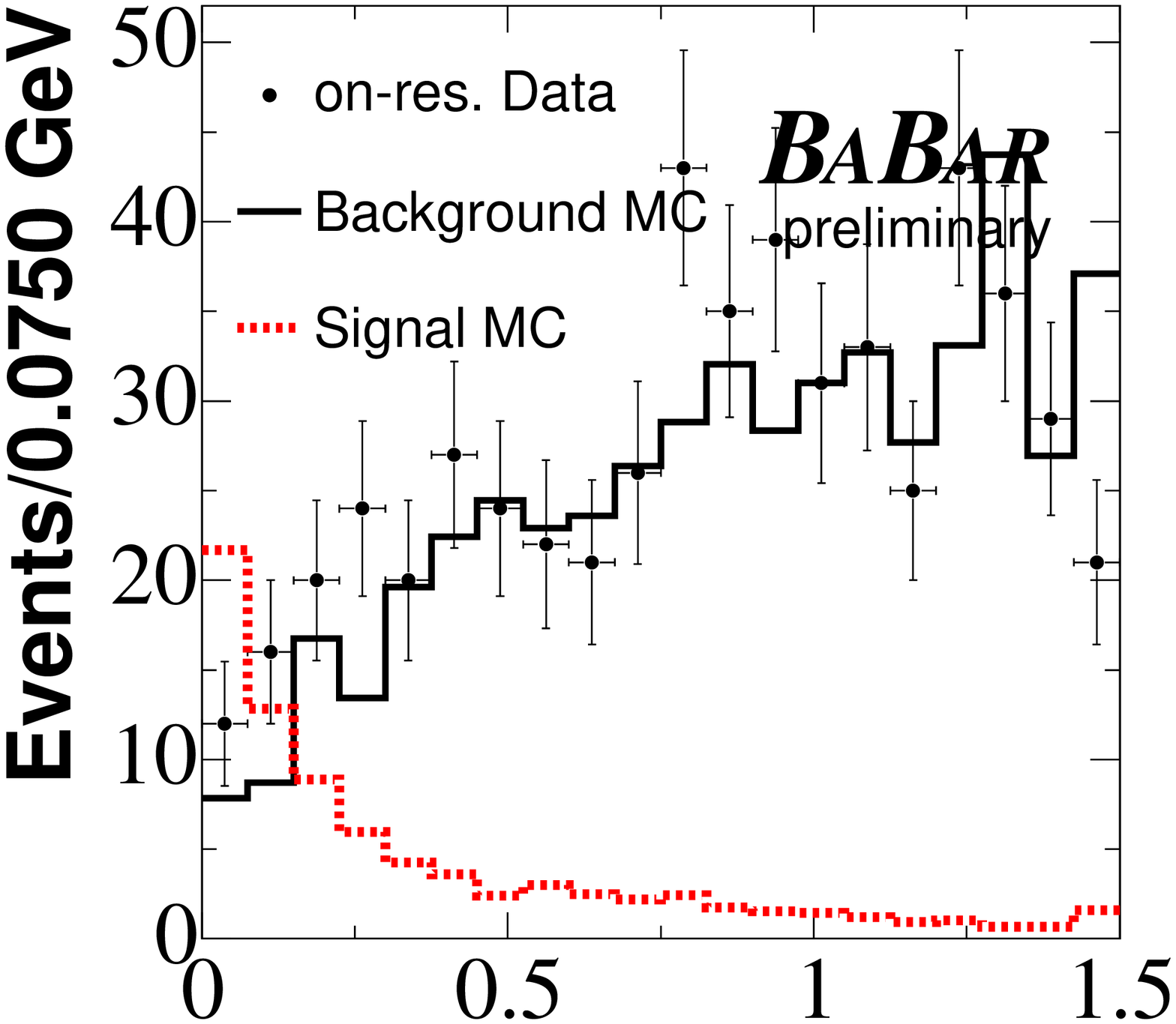}}}
     \hspace{.02in}
      \subfigure[The distribution of the $E_{\rm{extra}}$ 
     	for events satisfying $\tautopipiznu$ selection.]{
          \label{fig:EextraSB_pipi0nu_eCNSum}
          {\includegraphics[width=.31\textwidth]{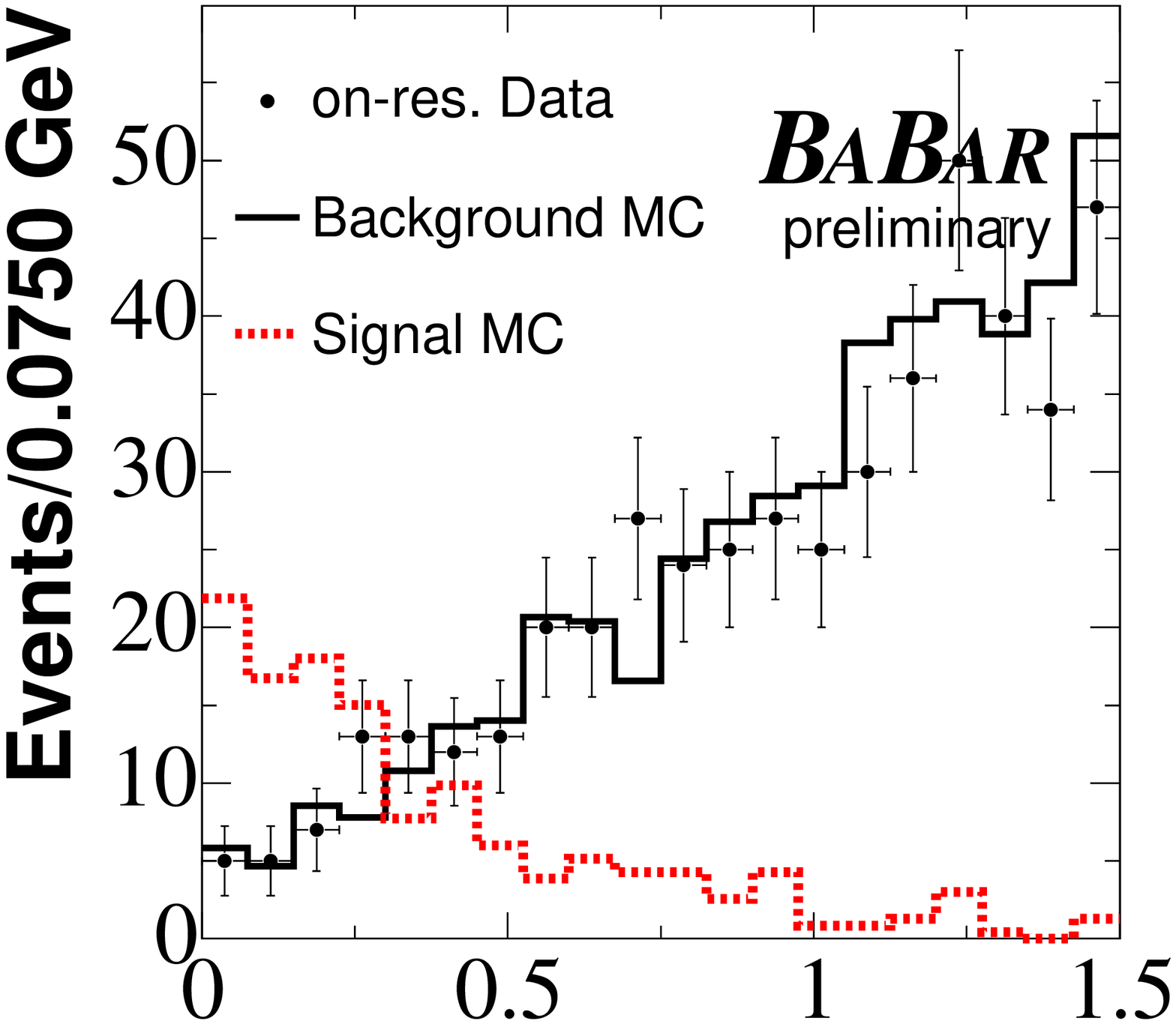}}}
     \hspace{.02in}
      \subfigure[The distribution of the $E_{\rm{extra}}$ 
     	for events satisfying $\tautothreepinu$ selection.]{
          \label{fig:EextraSB_3pinu_eCNSum}
          {\includegraphics[width=.31\textwidth]{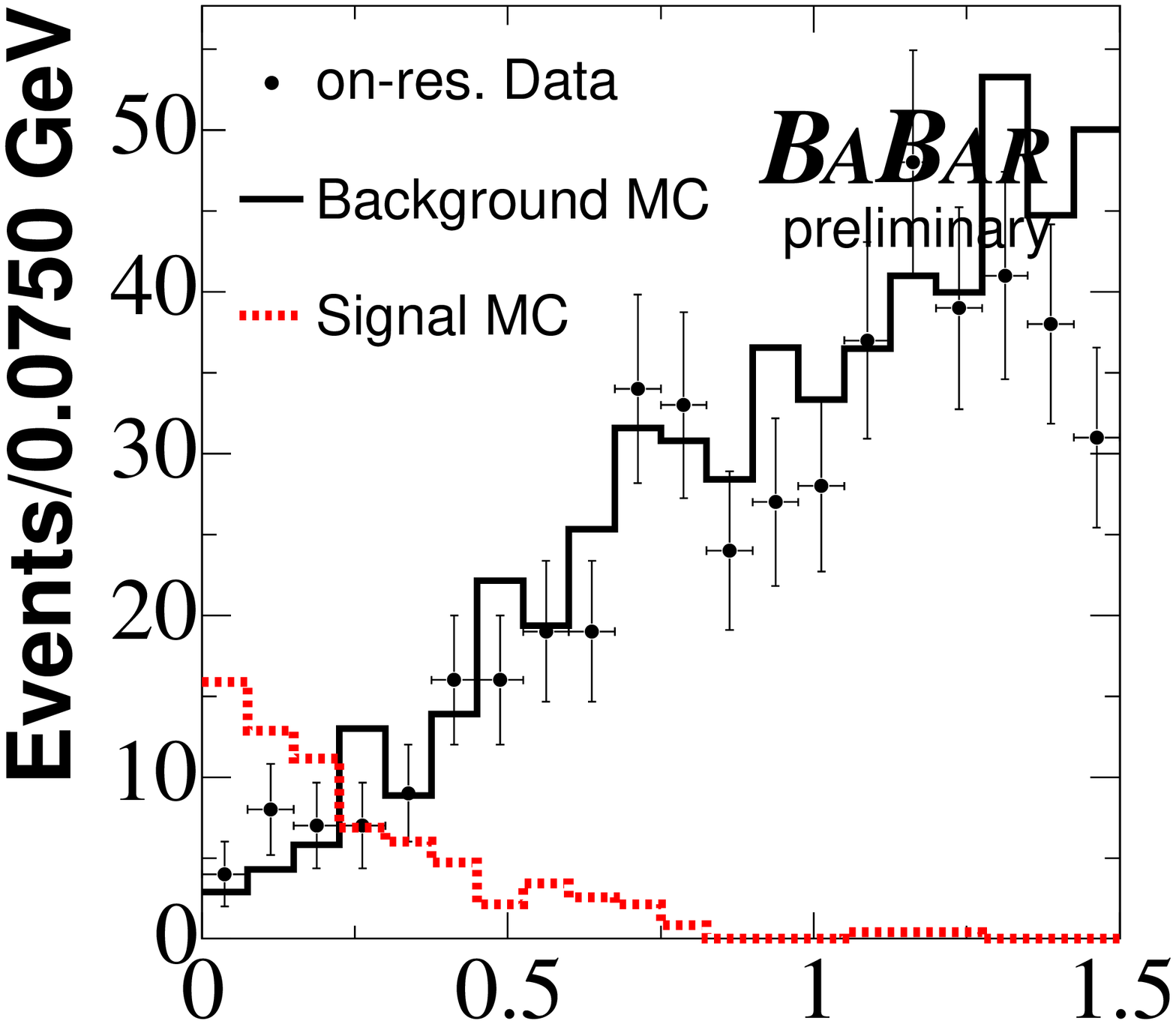}}}\\
     \caption{The distribution of the $E_{\rm{extra}}$ for events passing the selections for different $\tau$ decay modes, plotted for background and signal MC simulation and on-resonance data. Events plotted here are required to pass selections for corresponding signal modes. The background MC events are normalized to on-resonance data luminosity. The signal MC events for the $\tautoenunu$, $\tautomununu$, and $\tautopinu$ modes are normalized assuming $\mathcal{B}(\btn)=10^{-3}$, whereas for the normalization of the signal MC events in the $\tautopipiznu$ and $\tautothreepinu$ modes we use $\mathcal{B}(\btn)=10^{-2}$. 
The above $\mathcal{B}(\btn)$ assumptions for the normalization of the signal MC events are used to compare the shapes of the $E_{\rm{extra}}$ distributions between signal MC events, background MC simulation, and on-resonance data events. }
     \label{fig:EextraSB_0to1500mev_BkgOnly}
\end{figure}

After finalizing the signal selection criteria, the signal region
($E_{\rm{extra}} < 0.3 \gev$) in the on-resonance data is examined. 
Table \ref{tab:unblind-result} lists the
number of observed events in on-resonance data in the signal region,
together with the expected number of background events in the 
signal region.
Figure \ref{fig:EextraSB_0to1500mev_BkgOnly} shows the $E_{\rm{extra}}$ 
distribution in data and simulation. 
In almost all the modes the
observed number of events is in agreement with the expected 
number of background events.

\begin{table}[hbt]
\centering
\caption{\label{tab:unblind-result}Shown are the observed number of on-resonance data events in the signal region, together with number of expected background events. The background estimations include systematic corrections referred to in Sec. \ref{sec:bkgEstSys}.}
\vspace{0.2cm}
\begin{tabular}{|c|c|c|} \hline
Selection    & Expected           &  Observed Events  \\ 
             & Background Events  &  in On-resonance Data  \\ \hline
$\enunu$     & 15.15 $\pm$ 3.14   & 13  \\ \hline
$\mununu$    &  8.05 $\pm$ 2.07   & 10  \\ \hline
$\pinu$      & 55.30 $\pm$ 7.37   & 72  \\ \hline
$\pipiznu$   & 29.80 $\pm$ 5.10   & 30  \\ \hline
$\threepinu$ & 25.10 $\pm$ 3.87   & 26  \\ \hline
\hline
\end{tabular}
\end{table}

We determine the \btn \xspace branching fraction from the number of signal
candidates $s_i$ in data for each $\tau$ decay mode, according to $s_i =
N_{B \overline{B}}  \mathcal{B}(\btn) \varepsilon_{\rm{tag}} \varepsilon_i$. Here 
$N_{B \overline{B}}$ is the total number of $B\overline{B}$ pairs in data,
$\varepsilon_{\rm{tag}}$ is the tag reconstruction efficiency in signal 
MC; $\varepsilon_i$ is the signal-side selection efficiency in
different $\tau$ decay modes calculated with respect to the total number
of reconstructed tag B mesons. Table \ref{tab:EffBcountSummary} shows 
the values of $N_{B \overline{B}}$, $\varepsilon_{\rm{tag}}$ and 
$\varepsilon_i$, after applying appropriate systematic corrections 
(sec. \ref{sec:Systematics}).

\begin{table}[hbt]
\centering
\caption{\label{tab:EffBcountSummary}Shown are the values for the quantities  $N_{B \overline{B}}$, $\varepsilon_{\rm{tag}}$ and 
$\varepsilon_i$, after applying systematic corrections.}
\begin{tabular}{|c|c|c|} \hline
                        &          Value                               & Total Uncertainty (\%) \\
\hline \hline								     
$N_{B \overline{B}}$    &  $(124.1 \pm 1.4) \times 10^{6}$             &  1.1  \\ \hline 
$\varepsilon_{\rm{tag}}$     & $(1.82 \pm 0.074 \pm 0.055)\times 10^{-3}$   & 5.06  \\ \hline		
$\varepsilon_{\enunu}$                & (8.36  $\pm$  0.42  $\pm$  0.28) \% &  6.03  \\ \hline
$\varepsilon_{\mununu}$               & (4.30  $\pm$  0.28  $\pm$  0.17) \% & 7.56   \\ \hline
$\varepsilon_{\pinu}$                 & (22.34  $\pm$  0.72  $\pm$ 1.36) \% & 6.91   \\ \hline
$\varepsilon_{\pipiznu}$              & (3.01  $\pm$  0.24  $\pm$  0.25) \% & 11.45   \\ \hline
$\varepsilon_{\threepinu}$            & (2.07  $\pm$  0.20   $\pm$ 0.13) \% & 11.53   \\ \hline

\hline
\end{tabular}
\end{table}

The results from each decay mode are combined using the ratio 
$Q = {\cal L}(s+b)/{\cal L}(b)$,
where ${\cal L}(s+b)$ and ${\cal L}(b)$ are the
likelihood functions for signal plus background and background-only
hypotheses, respectively~\cite{bib:cls, bib:lepcls}:
\begin{equation}
  {\cal L}(s+b) \equiv
  \prod_{i=1}^{n_{ch}}\frac{e^{-(s_i+b_i)}(s_i+b_i)^{n_i}}{n_i!},
        \;
  {\cal L}(b)   \equiv
  \prod_{i=1}^{n_{ch}}\frac{e^{-b_i}b_i^{n_i}}{n_i!},
  \label{eq:lb}
\end{equation}

\begin{figure}[htp]
     \centering
     {\includegraphics[width=.6\textwidth]{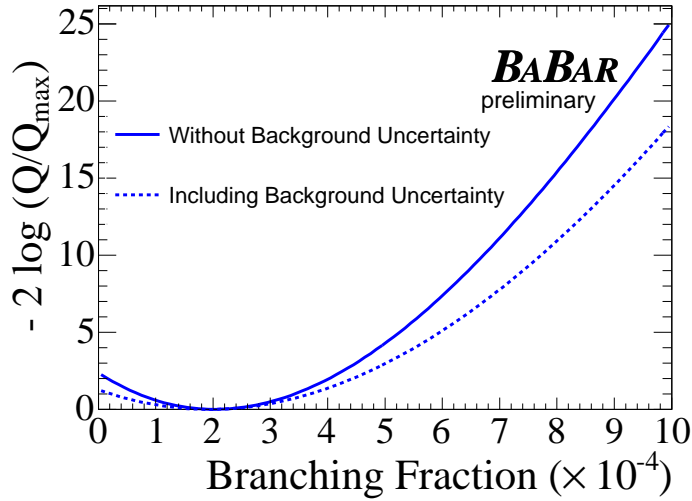}}
     \caption{The distribution of likelihood ratio as a function of $\mathcal{B}(\btn)$. The dashed (solid) curves corresponds to the case in which the uncertainty on the expected background is included (not included).}
     \label{fig:logQoverQmax}
\end{figure}

Since we have no evidence of signal we set an upper limit.
The statistical and systematic uncertainties on the expected
background ($b_{i}$) are included in the likelihood definition by
convolving it with a Gaussian distribution ($\mathcal{G}$).
The mean of $\mathcal{G}$ is $b_{i}$, and
the standard deviation ($\sigma_{b_{i}}$) of $\mathcal{G}$ is the  
statistical and systematic errors on $b_i$ added 
in quadrature~\cite{bib:lista},

\begin{equation}
\calL(s_i+b_i) \rightarrow \calL(s_i+b_i) \otimes \mathcal{G}(b_i,\sigma_{b_{i}})%
\;
\end{equation}
(similarly for $\calL(b_i)$).
We determine the limit on the branching fraction to be
$\mathcal{B}(\btn) < 4.3 \times 10^{-4}$ at the $90\%$ C.L.
Figure \ref{fig:logQoverQmax} shows the likelihood ratios as a function 
of $\mathcal{B}(\btn)$. The solid curve corresponds to the case 
in which the uncertainty on the expected background is not included.
The effect of including the uncertainty on the expected background
can be seen from the dashed curve.

The measured branching fraction, the value that maximizes the likelihood
ratio estimator is $1.9^{+1.8}_{-1.7} \times 10^{-4}$.

The \babar\ Collaboration performed also a search for the $\btn$ decay
\cite{babar_prl_btn}, where 
the tag B mesons are reconstructed in hadronic modes 
$B^{-} \to D^{(*)0} X^{-}$. Here $X^{-}$ represents a combination of up to 
five charged pions or kaons and up to two $\piz$ candidates. The
hadronic reconstruction analysis is statistically independent from the 
current analysis and has obtained a limit
$\mathcal{B}(\btn) < 4.2 \times 10^{-4}$ at the $90\%$ C.L.
We combine the results from the
statistically independent hadronic and semi-leptonic samples by first
calculating the likelihood ratio estimator,
$Q \equiv \calL(s+b)/\calL(b)$ using the likelihood functions
from each method. We then create a combined estimator from the product
of the semi-leptonic ($Q_{\rm{sl}}$) and hadronic ($Q_{\rm{had}}$) likelihood ratio
estimators, $Q = Q_{\rm{sl}} \times Q_{\rm{had}}$. The combined upper limit
on the branching fraction is 
$\mathcal{B}(\btn) < 3.3 \times 10^{-4}$ at the $90\%$ C.L.

Using equation \ref{eqn:br}, the combined 
branching fraction upper limit, and the measured value of $|V_{ub}|$ 
\cite{ref:pdg2004} we set a limit on $f_{B}$. We find
$f_{B} < 0.480$ \gev\ at 90\% C.L.

The \babar\ Collaboration also searched for \btn\ decays
using semi-exclusive semi-leptonic decay modes \btodlnux\,
where $\ell=e,\mu$ and $X$ can be a $\gamma$, \piz, or nothing
\cite{babar_prl_btn}.
In the signal-side only the leptonic $\tau$ decays, $\tautoenunu$
and $\tautomununu$, are identified. 
The \btodszlnu\ tags used in this current analysis overlap with 
the \btodlnux\ tags and the analyses are therefore not combined.

\section{SUMMARY}
\label{sec:Summary}

We have performed a search for the decay
process \btn. To accomplish this a sample of
semi-leptonic $B$ decays (\dszlnu) has been used to
reconstruct one of the $B$ mesons and the remaining information in
the event is searched for evidence of \btn. We find no evidence for this
decay process and set a preliminary limit on its branching fraction of
\begin{eqnarray*}
\mathcal{B}(\btn) & < & 4.3 \times 10^{-4} \, \textrm{at the 90\% CL.}
\end{eqnarray*}
By combining this analysis with a statistically independent
\btn\ search performed using a hadronic $B$ reconstruction
we find the preliminary combined limit:
 
\begin{eqnarray*}
\mathcal{B}(\btn)_{combined}< 3.3 \times10^{-4} \, \textrm{at the 90\% CL.}
\end{eqnarray*}

\section{ACKNOWLEDGMENTS}
\label{sec:Acknowledgments}
We are grateful for the 
extraordinary contributions of our \pep2\ colleagues in
achieving the excellent luminosity and machine conditions
that have made this work possible.
The success of this project also relies critically on the 
expertise and dedication of the computing organizations that 
support \babar.
The collaborating institutions wish to thank 
SLAC for its support and the kind hospitality extended to them. 
This work is supported by the
US Department of Energy
and National Science Foundation, the
Natural Sciences and Engineering Research Council (Canada),
Institute of High Energy Physics (China), the
Commissariat \`a l'Energie Atomique and
Institut National de Physique Nucl\'eaire et de Physique des Particules
(France), the
Bundesministerium f\"ur Bildung und Forschung and
Deutsche Forschungsgemeinschaft
(Germany), the
Istituto Nazionale di Fisica Nucleare (Italy),
the Foundation for Fundamental Research on Matter (The Netherlands),
the Research Council of Norway, the
Ministry of Science and Technology of the Russian Federation, and the
Particle Physics and Astronomy Research Council (United Kingdom). 
Individuals have received support from 
CONACyT (Mexico),
the A. P. Sloan Foundation, 
the Research Corporation,
and the Alexander von Humboldt Foundation.

\end{document}